\documentclass[journal ]{new-aiaa}
\usepackage[utf8]{inputenc}
\usepackage{textcomp}
\usepackage{bm}
\usepackage{graphicx}
\usepackage{amsmath}
\usepackage[version=4]{mhchem}
\usepackage{siunitx}
\usepackage{subcaption}
\usepackage{caption} 
\usepackage{multirow}
\usepackage{longtable,tabularx}
\title{Nonlinear Unsteady Vortex-Lattice Vortex-Particle Method with Adaptive Wake Conversion for Rotorcraft Aerodynamics}

\author{Jinbin Fu \footnote{Post-doctoral fellow, Department of Mechanical Engineering, Polytechnique Montréal, jinbin.fu@polymtl.ca, jinbin.fu@polimi.it.} and Eric Laurendeau\footnote{Professor, Department of Mechanical Engineering, Polytechnique Montréal, eric.laurendeau@polymtl.ca, AIAA Senior Member}}
\affil{Polytechnique Montréal, Montréal, Quebec, Canada, H3T 1J4}

\begin{document}

\maketitle

\begin{abstract}
Nonlinear unsteady vortex lattice–vortex particle methods (NL-UVLM-VPM) provide medium-fidelity predictions of rotorcraft aerodynamics with explicit three-dimensional wake representations at a moderate computational cost. This study presents an NL-UVLM-VPM approach with a scale-consistent adaptive wake panel-particle conversion strategy that mitigates the inherent temporal–spatial resolution coupling of conventional wake treatments in rotorcraft aerodynamic simulations. Numerical assessment shows that this strategy preserves the near third-order temporal convergence of the underlying time-integration scheme while improving robustness under coarsened temporal resolution. For a representative hover case, computational time is reduced by 29$\%$ relative to the conventional conversion strategy at identical temporal resolution and by nearly 70$\%$ compared with a fine-resolution reference simulation over 20 rotor revolutions, while maintaining thrust and torque predictions within 1$\%$ of the reference solution. Based on these analyzes, practical recommendations for particle conversion parameters and wake resolution are provided. The methodology is further validated for increasingly complex scenarios, including hover, forward flight with blade–vortex interaction, and multirotor interaction. Predictions show good agreement with experimental data and dedicated unsteady Reynolds-averaged Navier–Stokes simulations (URANS), while computational speedups exceeding two orders of magnitude relative to URANS are achieved.
\end{abstract}

\section*{Nomenclature}
{\renewcommand\arraystretch{1.0}
\noindent\begin{longtable*}{@{}l @{\quad=\quad} l@{}}
$\mathrm{C_D}$ & airfoil drag coefficient, $D/(1/2)\rho_{\infty} U_{\infty}^2 c$ \\
$\mathrm{C_L}$ & airfoil lift coefficient, $L/(1/2)\rho_{\infty} U_{\infty}^2 c$ \\
$\mathrm{C_P}$& sectional pressure coefficient, $P/(1/2)\rho_{\infty} U_{\infty}^2$ \\
$\mathrm{C_Q}$ & torque coefficient, $Q/\rho_{\infty} U^2 \pi R^3$ \\
$\mathrm{C_T}$ & thrust coefficient, $T/\rho_{\infty} U^2 \pi R^2$ \\
$\mathrm{c}$ & chord length, m \\
$c_V$ & the standard Vreman subgrid-scale model coefficient \\
$\mathrm{d\mathbf{l}}$ & vector length of the straight-line vortex segment, m \\
$\mathrm{M}$ & Mach number \\
$\mathrm{N_{rev}}$ & number of rotor revolutions \\
$\mathrm{N_b}$  & number of blades per rotor \\
$\mathrm{N_p}$  & number of vortex particles \\
${\mathrm{N_{\text{steps}}}}$  & {number of time steps per rotor revolution} \\
$\mathrm{n_c}$  & number of panel chordwise discretization  \\
$\mathrm{n_i}$  & number of particles assigned at the $i$-th straight-line element \\
$\mathrm{n_s}$  & number of panel spanwise discretization \\
$\mathrm{n_t}$  & number of particles assigned at the tip vortex segment \\
$\mathrm{Q}$ & rotor torque, $\mathrm{N \cdot m}$ \\
$\mathrm{R}$ & rotor radius, m \\
$\mathrm{R_c}$ & cut-out rotor radius, m \\
$\mathrm{Re}$ & Reynolds number \\
$\mathrm{r}$ & radius position, m \\
$\mathrm{T}$ & rotor thrust, N \\
$\mathrm{t}$ & time, s \\
$\mathrm{\bm{u}}$ & velocity, m/s \\
$\mathrm{\mathbf{x}_i}$ & the position of the $i$-th particle, $\mathrm{\mathbf{x}_i}(t)$, m \\
$\alpha$ & angle of attack, $\mathrm{\deg}$ \\
$\mathrm{\psi}$ & azimuthal angle, $\mathrm{\deg}$ \\
$\theta$ & blade pitch angle, $\mathrm{\deg}$ \\
$\beta$  & blade flap angle, $\mathrm{\deg}$ \\
$\delta$  & blade lead-lag angle, $\mathrm{\deg}$ \\
$\mathrm{\bm{\Gamma}_i}$ & vortex strength of $i$ th particle, $\mathrm{\bm{\Gamma}_i}(t)$, $\mathrm{m^3/s}$ \\
$\rho$ & 3D Gaussian smoothing function parameter \\
$\rho_{\infty}$ & air density, $\mathrm{kg/m^3}$\\
$\sigma$  & filter width (smoothing radius or core size), $\mathrm{m}$ \\
$\zeta_{\sigma}$ & 3D Gaussian distribution, $\mathrm{1/m^2}$ \\
$\nu$ & kinematic viscosity, $\mathrm{m^2/s}$ \\
$\mathrm{\bm{\omega}}$ & vorticity field, $\mathrm{\bm{\omega}(\mathbf{x},t)}$, $\mathrm{1/s}$\\ 
$\Delta x_{\text{tip}}$ & blade tip particle spacing, $\mathrm{deg}$ \\
$\Delta \Psi$ & blade azimuthal increment, $\mathrm{deg}$ \\
\multicolumn{2}{@{}l}{Subscripts} \\
extrap & Richardson extrapolation  \\
\end{longtable*}}
\addtocounter{table}{-1}

\section{Introduction}
\lettrine{M}{odeling} rotorcraft aerodynamics accurately and efficiently remain significant challenges due to the highly complex aerodynamic environment inherent to rotor systems. With the increasing interest in urban air mobility (UAM), there is a growing demand for aerodynamic tools that balance fidelity and computational cost. High-fidelity approaches such as three-dimensional(3D) unsteady Reynolds-Averaged Navier-Stokes (URANS) and Scale-Resolving Simulations (SRS) can capture detailed flow physics but require fine spatial resolution to mitigate numerical dissipation and remain computationally expensive for long-duration simulations.\cite{Fu2023.1, Chaderjian2012, Narducci2015, Jain2015}.

To reduce computational cost while retaining essential wake physics, 
several classes of mid-fidelity aerodynamic tools for rotor analysis have been developed. Hybrid approaches combining potential-flow blade models with free-wake or vortex particle methods (VPM) are widely used \cite{Yin2000, Opoku2002, Zhao2010, Tan2013, Lee2019, Tugnoli2021, Yurt2025}. Representative implementations include the DLR unsteady panel method (UPM) \cite{Yin2000} and DUST software \cite{Tugnoli2021}. In parallel, particle-based rotor formulations employing actuator-line or actuator-surface representations have also been explored, as exemplified by FlowUnsteady \cite{Alvarez2024}. Collectively, these methods enable rotor performance prediction at substantially lower computational costs than high-fidelity grid-based CFD while preserving dominant wake structures.

Within this context, the authors' research group developed a hybrid nonlinear unsteady vortex lattice-vortex particle method (NL-UVLM-VPM) incorporating a viscous-inviscid $\alpha$-coupling strategy supported by a two-dimensional (2D) RANS database and a Particle Strength Exchange (PSE)-based viscous diffusion model within a Large Eddy Simulation (LES) framework. As reported in Ref.~\cite{Proulx-Cabana2022}, the method achieved good correlation with experimental data for the hover performance of the S-76 rotor model. However, that validation was limited to steady performance prediction, and the predictive capability of the approach in more complex unsteady aerodynamic environments, such as forward flight and multirotor interaction, remains insufficiently validated. These operating conditions introduce additional challenges, including wake skewing, blade-vortex interaction, and rotor-rotor interaction, which require further systematic assessment.

In conventional hybrid UVLM-VPM implementations, wake particles are shed at every time step, inherently coupling temporal and spatial discretizations, which introduces a trade-off between computational efficiency, spatial resolution, and numerical robustness \cite{Yurt2025}. Relaxing temporal resolution to improve efficiency reduces wake resolution and degrades predictive accuracy, whereas increasing particle density to maintain spatial resolution can compromise numerical robustness. As a result, this spatio-temporal coupling limits the resolution flexibility and numerical robustness of long-duration simulations.

Building upon the current NL-UVLM-VPM solver, the present work introduces a scale-consistent adaptive wake panel-particle conversion strategy based on the geometric streamwise length of trailing vortex segments. By assigning particle density proportional to the actual wake-segment arc length, this method improves resolution consistency along the rotor wake, thereby mitigating the inherent temporal-spatial discretization trade-off.

The objectives of this study are threefold. First, to introduce an adaptive wake panel-particle conversion strategy within the NL-UVLM-VPM framework. Second, to assess its numerical convergence, robustness, and algorithmic efficiency in representative hover simulations, and to provide practical recommendations for numerical stability and efficiency. Third, to extend the validation envelope of the solver beyond hover and evaluate its predictive capability and computational cost under more complex operating conditions, including forward flight and rotor-rotor interaction, through comparisons with dedicated URANS simulations and available experimental data.

The remainder of this paper is organized as follows. Section~\ref{sec2} summarizes the NL-UVLM and VPM formulations and presents the adaptive wake panel-to-particle conversion strategy. Section~\ref{sec3} examines the numerical convergence, robustness, and algorithmic efficiency of the adaptive conversion method, including a temporal-resolution sensitivity study and recommended parameter settings. Section~\ref{sec4} provides validation results for benchmark rotor configurations, including hover, forward flight, and rotor-rotor interaction cases, with comparisons against URANS simulations and experimental measurements, together with an assessment of overall computational cost. Finally, the main conclusions are summarized in {Section~\ref{sec5}.

\section{Numerical Methods} \label{sec2}

\subsection{Nonlinear unsteady vortex-lattice method (NL-UVLM)} \label{subsec1}
The unsteady vortex-lattice method (UVLM) is a low-order computational technique for solving unsteady potential flow around lifting surfaces. A detailed description of this method can be found in Ref.~\cite{Proulx-Cabana2022}. Here, a summary is presented for completeness and to help understand the additional developments presented later in the paper.

As shown in Fig.~\ref{fig1}, in rotor simulations, the blade geometry is treated as a thin body composed of lifting surfaces, and the wake is represented as a set of vortex panels. The blade and wake are simulated using the classical vortex-lattice method \cite{Katz2001} with time-stepping, referred to as the unsteady vortex-lattice method (UVLM). Specifically, the blade planform is simplified to a flat surface between the leading and trailing edges, discretized using a structured grid. Vortex ring elements are placed at the quarter-chord location of each grid element. For the wake, vortex panels extend from the trailing edge to infinity, with the Kutta condition applied to ensure that the circulation equals zero. 

\begin{figure}[h]
    \centering
    \includegraphics[width=0.8\columnwidth]{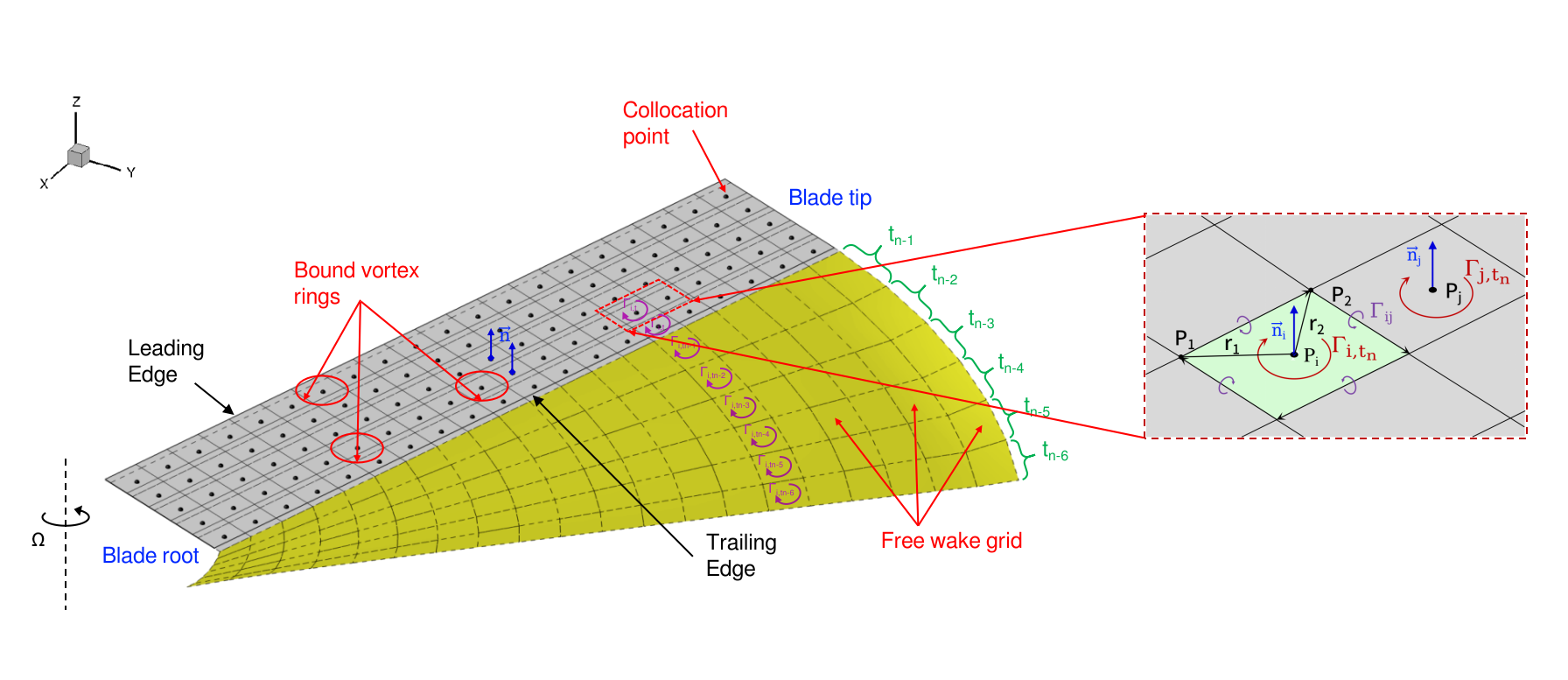}
    \caption{Descriptive diagram of the Unsteady Vortex-Lattice Method (UVLM).}
    \label{fig1}
\end{figure}

The induced velocity from each vortex ring element is computed using the Biot-Savart law, with the Vatistas smoothing kernel \cite{Vatistas1991} to eliminate the singularity in the Biot-Savart equation during wake displacement in the rotor simulations. A Neumann boundary condition, which enforces the local velocity to be tangent to the vortex ring element, is applied to the collocation points. These points are located at the center of the three-quarter-chord line of each lifting surface panel. This treatment ensures that the vortex rings generate lift with a slope consistent with thin airfoil theory.

To include nonlinear effects, a low computational cost viscous-inviscid $\alpha$-iterative coupling algorithm is applied to each chordwise strip of vortex panels along the span of the blade. Each strip is corrected using a 2D viscous database formatted as a .c81 table, which has been calculated with CHAMPS \cite{Parenteau2020}. The iterative viscous coupling is detailed in Proulx-Cabana's work \cite{Proulx-Cabana2022}.

Aerodynamic loads are determined by integrating the UVLM solution with the viscous section data, taking advantage of the availability of both information. The steady component of the loads is computed on each panel side from the Kutta-Joukowski theorem \cite{Simpson2013}, while the unsteady component is evaluated at the center of every panel through the unsteady Bernoulli equation \cite{Simpson2013}.

\subsection{Vortex particle method (VPM)}
The vortex particle method is a Lagrangian, grid-free approach that models the evolution of free wakes using viscous vortex particles to construct the vorticity field governed by the velocity-vorticity formulation of the incompressible Navier-Stokes equations:
\begin{equation}\label{eq1}
\frac{D\bm{\omega}}{Dt} = (\bm{\omega} \cdot \nabla) \mathbf{u} + \nu \nabla^2 \bm{\omega}
\end{equation}
where $\bm{\omega} = \nabla \times \mathbf{u}$ is the vorticity field associated with the velocity field, and $\nu$ is the kinematic viscosity of the flow. The first term on the right-hand side of Eq.~\eqref{eq1} is the vortex stretching term that controls the stretching and deformation of the vortex, while the second term represents the viscous diffusion effects.

In VPM, the vorticity field can be expressed in a Lagrangian manner as the sum of the contributions of a set of $N_p$ vector-valued particles, $i_p$:
\begin{equation}\label{eq2}
\bm{\omega}(\mathbf{x}, t) = \sum_{i=1}^{N_p} \bm{\Gamma_i}(t) \zeta_\sigma (\mathbf{x} - \mathbf{x}_{i}(t))
\end{equation} 
where $\mathbf{x}_{i}(t)$ and $\bm{\Gamma}_{i}(t)$ are the position and vorticity intensity, respectively, of the $i$-th particle. $\zeta_\sigma(\mathbf{x})$ is a smooth function, where $\sigma$ is the smoothing radius that is greater than the distance between particles to guaranty the convergence of VPM \cite{Beale1986}. In this work, the 3D Gaussian distribution is adopted as the smooth function:
\begin{equation}\label{eq3}
\zeta_\sigma(\rho) = \frac{1}{(2\pi \sigma^2)^{3/2}} e^{-{\rho^2}/{2}}
~~\mathrm{and}~~
\rho = \frac{|\mathbf{x} - \mathbf{x}_{i}(t)|}{\sigma}
\end{equation}
the parameter $\sigma$ is defined as the maximum distance from all neighboring particles in both the streamwise and spanwise directions, and it remains constant throughout the simulation.

Based on Eq.~\eqref{eq1}, the evolution of viscous vortex particles can be represented by the system of ordinary differential equations in convection-diffusion form,
\begin{equation}\label{eq4}
\frac{d \mathbf{x}_{i}(t)}{dt} = \mathbf{u}_{i}(t)
\end{equation}
\begin{equation}\label{eq5}
\frac{d \bm{\Gamma}_{i}(t)}{dt} = (\bm{\Gamma}_{i}(t) \cdot \nabla^\mathrm{T}) \mathbf{u}_{i}(t) + \nu \nabla^2 \bm{\Gamma}_{i}(t)
\end{equation}
where $i = 1, \ldots, N_p$ for all particles. The first equation, which describes the convection of the $i$-th vortex particle transported by the local flow velocity, is adopted to update the particle positions. The second equation expresses the effect of vortex stretching and diffusion on the vorticity intensity of the $i$-th vortex particle. In this work, both the particle positions and the vorticity intensity are advanced in time using the low-storage third-order Runge-Kutta scheme \cite{Williamson1980}. A Cartesian fast multipole method (FMM) is used to accelerate the calculation of the induced velocity and its gradient for vortex particles, reducing the numerical complexity from $O(N^2)$ to $O(N)$ \cite{Proulx-Cabana2022}.

The viscous diffusion term, which accounts for the effect of air viscosity on vorticity transportation, is solved using the PSE algorithm. To mitigate the instability of vortex stretching that may arise from insufficient viscous dissipation at subgrid scales~--~caused by the incomplete resolution of the full turbulent energy cascade described by Kolmogorov's theory \cite{Kolmogorov1991}~--~the Vreman subgrid scale (SGS) model \cite{Vreman2004} is adopted to compute turbulent eddy viscosity. This approach allows for modeling eddies at smaller scales and capturing coherent structures at coarser spatial resolutions than those required for direct numerical simulation (DNS). Further details can be found in a previous paper \cite{Proulx-Cabana2022}. In the present work, the Vreman model coefficient $C_V$ is set to 0.07, consistent with the value used in Ref.~\cite{Proulx-Cabana2022}.

Fig.~\ref{fig2} presents a DJI Phantom II rotor wake from the current VPM simulation at an advance ratio of $J = 0.35$. The simulation is able to capture every stage of the wake, from a coherent vortex structure to turbulent breakdown and mixing, and it predicts transitional mechanisms in between, such as short-wave instability, leapfrog, and pairing.
\begin{figure}[ht!]
\centering
\includegraphics[width=1.0\textwidth]{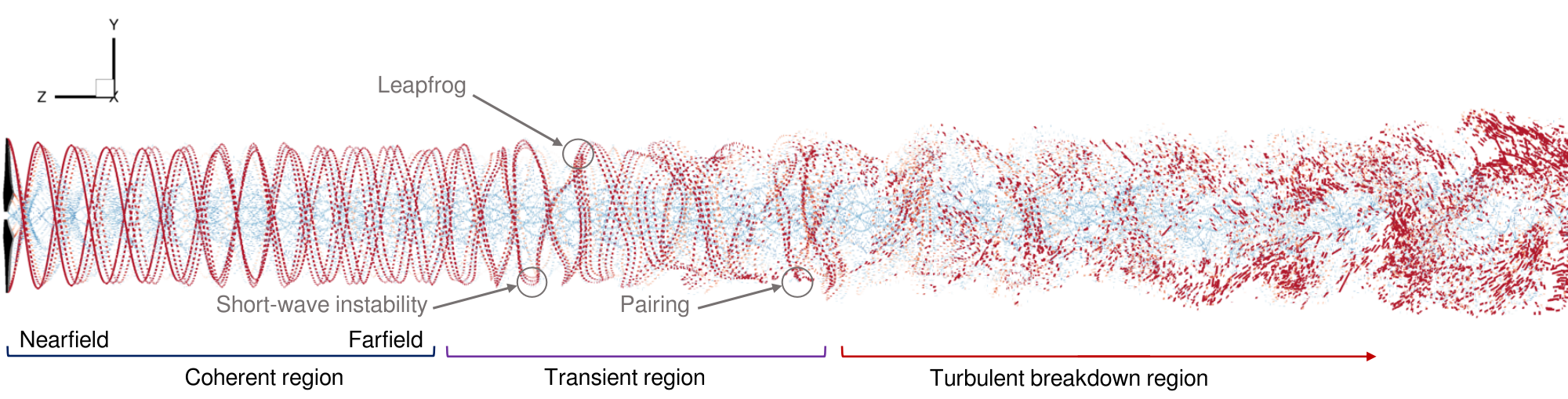}
\caption{The current VPM simulation capturing the wake evolution of a DJI Phantom II rotor at $J = 0.35$ from a coherent vortical structure to turbulent breakdown and mixing. \label{fig2}}
\end{figure} 

\subsection{Adaptive wake panel-particle conversion strategy}
In hybrid UVLM-VPM approaches, wake panels are commonly converted into vortex particles after a prescribed number of time steps to prevent excessive geometric distortion and numerical instability associated with panel stretching. Following this approach, the wake shed from the blade trailing edge is initially represented as straight-line vortex elements, with the strength of upstream vortex lines remaining unchanged to prevent circulation strength discontinuity at the trailing edge. These wake panels are then converted into vortex particles after a prescribed number of time steps, as illustrated in Fig.~\ref{fig3}. Here, the streamwise particles (shown in green) represent trailed particles, and the spanwise particles (shown in red) are shed particles. The strength of each new particle is proportional to the circulation difference between two adjacent wake panels, as given by:
\begin{equation} \label{eq6}
    \bm{\Gamma}_\text{i} = \frac{\Delta \Gamma_\text{i} d \mathbf{l}}{n_\text{i}}
\end{equation}
where $d \mathbf{l}$ is the vector length of the straight-line vortex element, {$n_\text{i}$ is the number of particles approximating the $i$-th straight-line element, and $\Delta\Gamma_\text{i}$ represents the circulation difference between two wake panels on either side of the edge. It can be defined as:
\begin{equation} \label{eq7}
\Delta\Gamma_\text{i} =  
\begin{cases}
\Gamma_\text{i}-\Gamma_{\text{i}_n}, &\mathrm{if~neighbour}~{i_\text{n}}~\mathrm{present}\\
\Gamma_\text{i}, &\mathrm{if~neighbour}~{i_\text{n}}~\mathrm{absent}
\end{cases}
\end{equation}
\begin{figure}[ht!]
\centering
\includegraphics[width=0.95\textwidth]{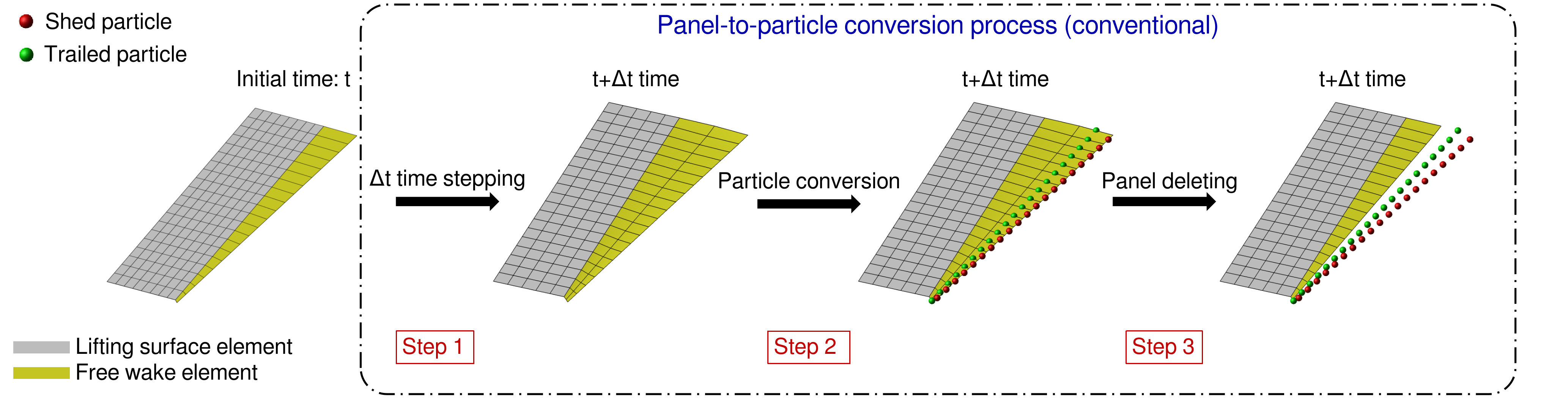}
\caption{The conventional conversion process of wake panels to particles ($n_\text{i}$ = 1)}. \label{fig3}
\end{figure} 

In the conventional wake panel-particle conversion strategy, each trailing vortex segment is discretized into an identical number of particles, regardless of its streamwise length. In rotor simulations, however, the streamwise length of trailing vortex segments increases with radial position, resulting in substantial variations in wake-segment geometry along the blade span. When uniform particle conversion is applied, this geometric inconsistency produces highly non-uniform spatial resolution, inefficient particle distribution, and degraded computational performance.

To improve resolution consistency, the present work introduces a scale-consistent adaptive wake panel-particle conversion strategy, in which the number of particles assigned to each trailing vortex segment is proportional to its actual streamwise arc length based on the prescribed particle count at the blade tip vortex segment. Let $\Delta s_\text{i}$ denote the streamwise length of the $\text{i}$-th trailing vortex segment and $\Delta s_\text{t}$ the reference length at the blade tip. Prescribing $n_\text{t}$ particles for the tip segment, the number of particles assigned to segment $\text{i}$ is defined as:
\begin{equation} \label{eq8}
    n_\text{i}=\left\lceil\frac{\Delta s_\text{i}}{\Delta s_\text{t}} \cdot n_\text{t}\right\rceil
\end{equation}
where $\left\lceil \cdot \right\rceil$ denotes upward rounding. Fig.~\ref{fig4} illustrates the resulting particle distribution obtained using the adaptive conversion method ($n_\text{t} = 2$). For clarity, the shed particles are omitted to highlight the conversion procedure. It can be observed that this adaptive scaling significantly improves spatial resolution consistency along the wake.
\begin{figure}[ht!]
\centering
\includegraphics[width=0.5\textwidth]{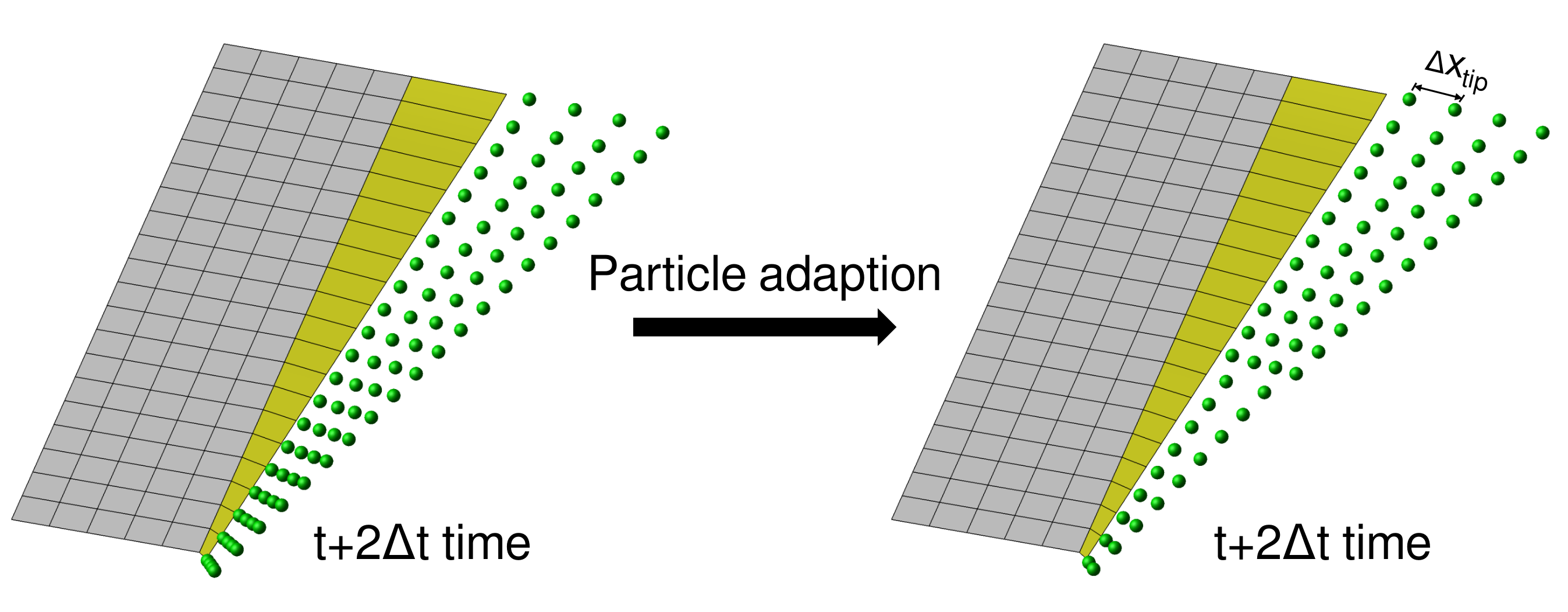}
\caption{The adaptive conversion process of wake panels to particles ($n_\text{t} = 2$) after two time steps. \label{fig4}}
\end{figure} 

\section{Numerical Convergence, Robustness and Efficiency of the Adaptive Conversion Strategy} \label{sec3}

\subsection{Resolution definition and decoupled parameterization}
To systematically assess the influence of the adaptive wake panel-particle conversion strategy on the temporal-spatial discretization behavior of the NL-UVLM-VPM solver, independent parameters are introduced to control temporal and wake spatial resolution. The temporal resolution is defined by the number of time steps per revolution, $N_\text{steps}$, from which the blade azimuthal increment is obtained as
\begin{equation} \label{eq9}
    \Delta \Psi = \frac{360^{\circ}}{N_\text{steps}}
\end{equation}
The wake spatial resolution is independently controlled through the prescribed tip particle spacing,
\begin{equation} \label{eq10}
    \Delta x_\text{tip} = \frac{\Delta \Psi}{n_\text{t}}
\end{equation}
where $n_\text{t}$ denotes the number of particles assigned to the blade-tip vortex segment. This parameterization enables variation in temporal resolution while maintaining constant wake spatial fidelity, therefore allowing direct assessment of the decoupling behavior introduced by the adaptive conversion strategy.

\subsection{Convergence behavior}
The convergence properties of the adaptive conversion strategy are evaluated using the Caradonna-Tung rotor in hover ($M_\text{T} = 0.439$, $\theta_\text{c} = 5^{\circ}$). Default parameters $N_{\text{steps}} = 144$, $\Delta x_{\text{tip}} = 2.5^{\circ}$, $n_\text{s} = 20$, and $n_\text{c} = 8$ are adopted as following the convergence studies reported in prior work \cite{Proulx-Cabana2022}, and serve as the reference case. In the present work, $\Delta \Psi$ is varied independently while maintaining the $\Delta x_{\text{tip}}$ constant, so as to evaluate the convergence behavior of the adaptive conversion algorithm without the influence of wake spatial refinement, and the thrust coefficient, $C_\text{T}$, is employed as the convergence metric.

Fig.~\ref{fig5} shows the convergence properties of the thrust coefficient, $C_{\text{T}}$, as the temporal resolution is refined by increasing the number of time steps per revolution, $N_{\text{steps}}$, from $9$ to $144$, corresponding to blade azimuthal increments of $40^{\circ}$ and $2.5^{\circ}$, respectively. Over this refinement range, the number of tip particles generated per revolution varies from $n_{\text{t}} = 16$ to $n_{\text{t}} = 1$. The corresponding Richardson extrapolation is also shown. An approximation of the convergence order, $p$, is obtained by fitting the numerical results to the model
\begin{equation} \label{eq11}
    C_{\text{T}} = (C_{\text{T}})_{\text{extrap}} + a(\Delta t)^p,
\end{equation}
where $\Delta t \propto 1/N_{\text{steps}}$, and the fitted curve is plotted as a solid line in the figure. The resulting convergence order is $p = 2.83$. This near third-order convergence confirms that the adaptive conversion strategy does not degrade the temporal accuracy of the third-order Runge-Kutta time integration used in the VPM solver. For $N_{\text{steps}} \geq 72$, the relative deviation in $C_{\text{T}}$ remains below $1\%$ with respect to the reference solution.
\begin{figure}[ht!]
\centering
\includegraphics[width=0.35\textwidth]{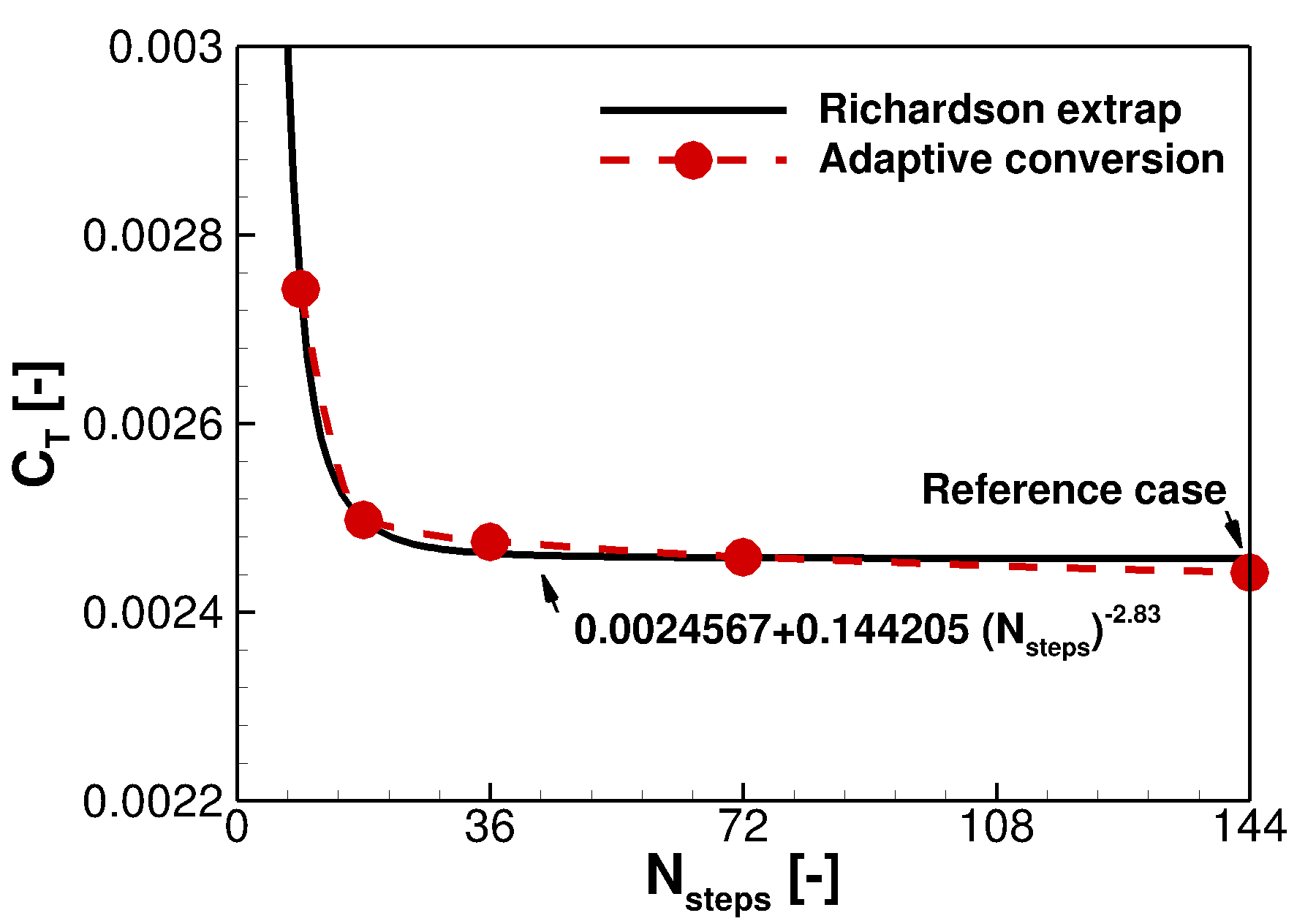}
\caption{The adaption conversion convergence study of the helicopter rotor simulation. \label{fig5}}
\end{figure}

\subsection{Numerical robustness under relaxed temporal resolution}
Fig.~\ref{fig6} compares the long-term behavior of the conventional and adaptive conversion strategies for coarse temporal discretizations. For $N_\text{steps} = 18$ to $36$, the conventional conversion exhibits numerical instability after approximately 8 and 10 rotor revolutions, manifested by divergence in $C_\text{T}$. In contrast, the adaptive conversion maintains stable solutions for 20 rotor revolutions across all tested temporal resolutions. The improved robustness is attributed to geometry-consistent scaling of particle density, which mitigates excessive near-root particle clustering and suppresses the growth of spurious numerical oscillations under relaxed temporal discretization. These results demonstrate that the adaptive strategy effectively alleviates the inherent temporal–spatial discretization trade-off present in the conventional conversion strategy.

\begin{figure*}[hbt!]
\centering
\begin{subfigure}{0.35\textwidth}
\includegraphics[width=\linewidth]{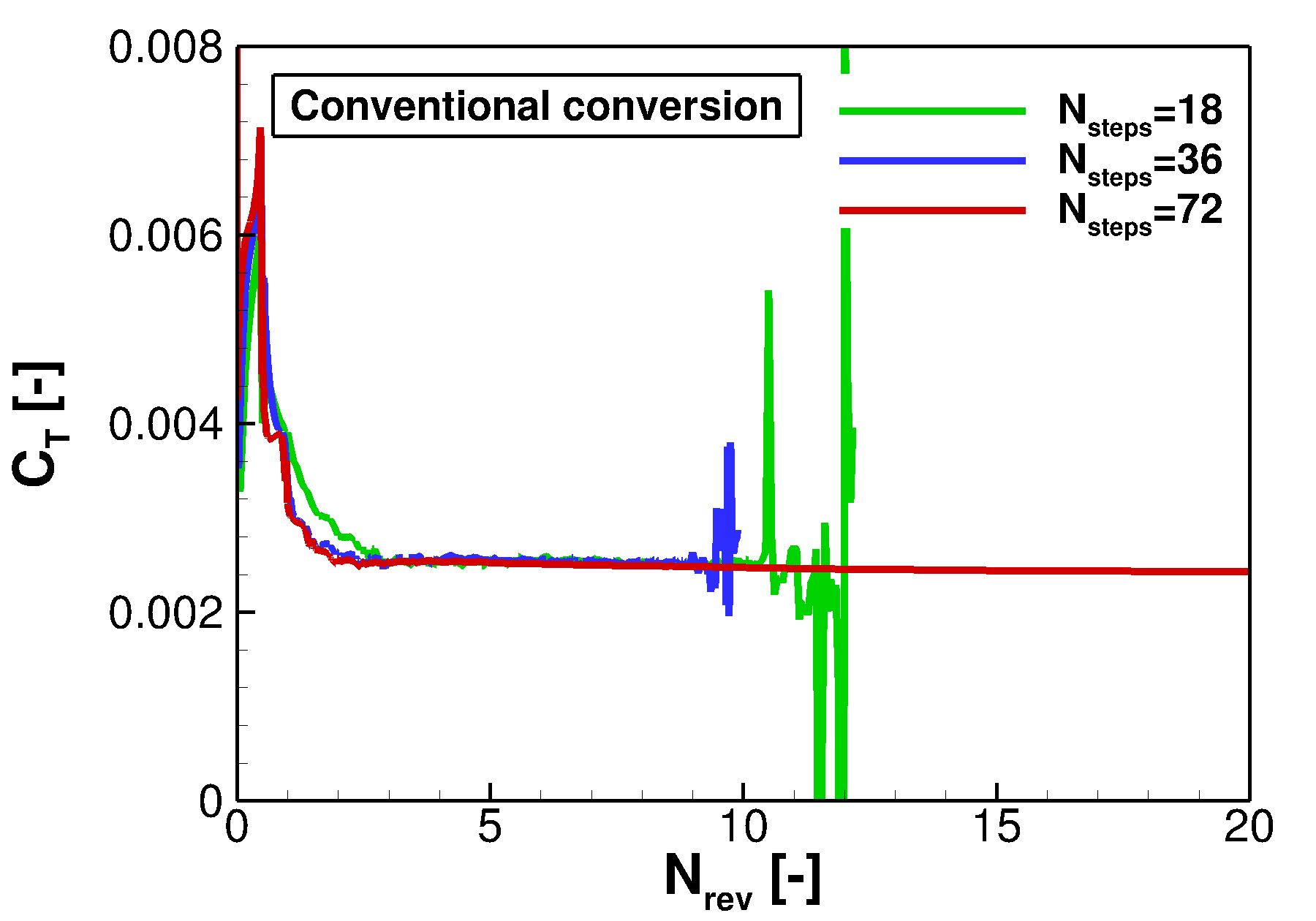}
\caption{$C_{\text{T}}$ for conventional conversion}
\end{subfigure}
\hspace{5.0mm}
\begin{subfigure}{0.35\textwidth}
\includegraphics[width=\linewidth]{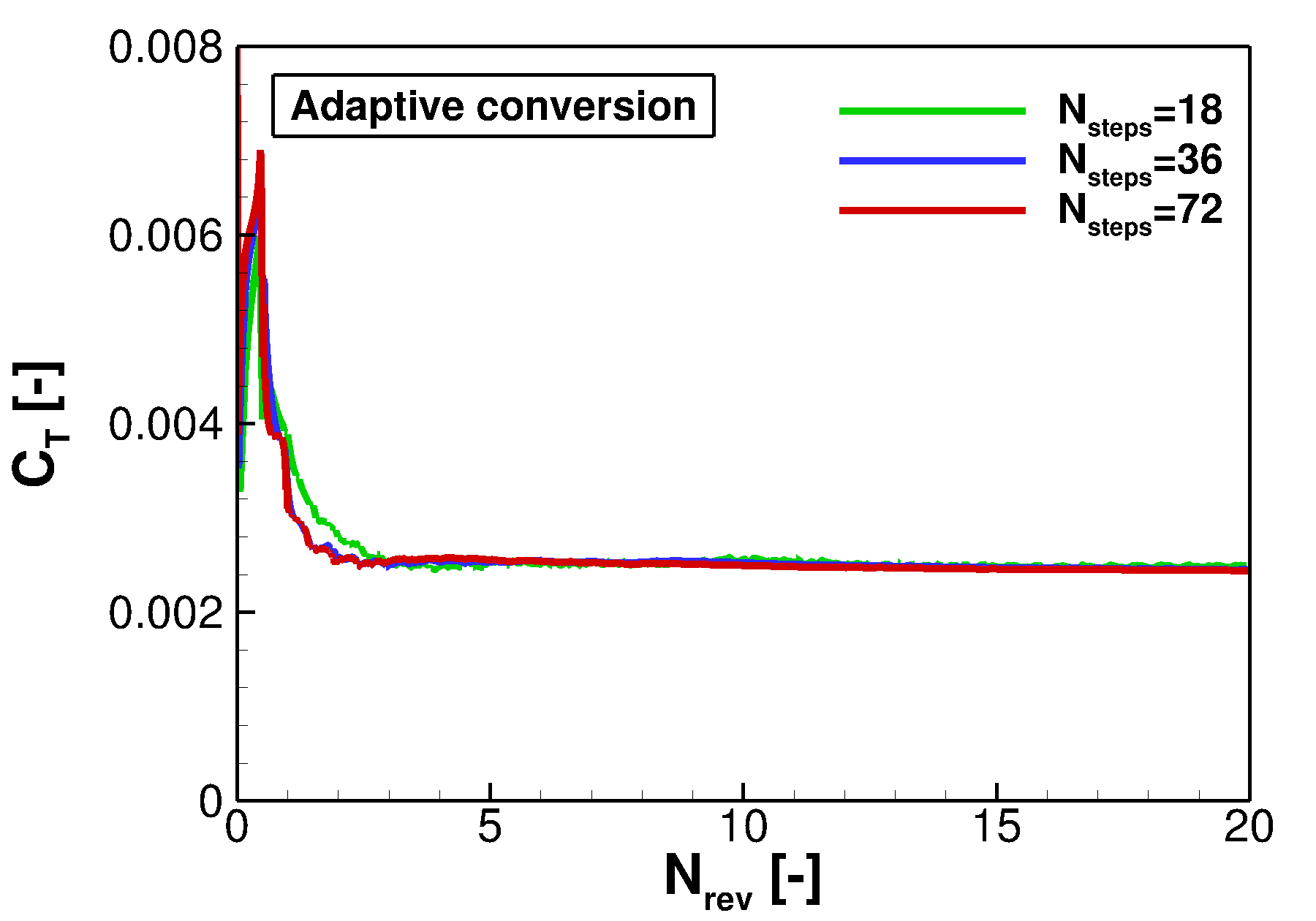}
\caption{$C_{\text{T}}$ for adaptive conversion}
\end{subfigure}
\caption{Convergence comparison of wake panel-particle conversion strategies.} \label{fig6} 
\end{figure*}

\subsection{Algorithmic efficiency and practical recommendation}
Table \ref{table1} provides quantitative comparisons of particle count and wall-clock time. All computations were conducted on a single AMD EPYC 9655 (Zen 5)@2.6 GHz processor with 96 cores and 192 threads for 20 rotor revolutions.
\begin{table}[hbt!]
\caption{Quantitative assessment of panel-particle conversion strategies.}
\centering
\begin{tabular}{lcccc}
\hline
Case & $\mathrm{C_T}$ deviation & $\mathrm{C_Q}$ deviation &Particles count & CPU wall-time \\\hline
Reference ($N_{\text{steps}}= 144$, $n_{\text{t}} = 1$)  & ------ & ------ & 235996 & 8.0 h \\
Conventional  ~~($N_{\text{steps}}= 72$,~~~$n_{\text{t}} = 2$)& $+0.7\%$ & $+0.09\%$ & 178312 & 3.5 h \\
Adaptive  ~~($N_{\text{steps}}= 72$,~~~$n_{\text{t}} = 2$) & $+0.6\%$ & $+0.09\%$ & 146676 & 2.5 h \\
\hline
\end{tabular} \label{table1}
\end{table}

At identical temporal resolution ($N_\text{steps} = 72$), the adaptive strategy reduces particle count by approximately $18\%$ and wall time by about $29\%$ relative to the conventional conversion, while maintaining comparable deviations in $C_{\text{T}}$ and $C_{\text{Q}}$. When compared with the fine-resolution reference case ($N_\text{steps} = 144$), the adaptive strategy enables an overall reduction of approximately $38\%$ in particle count and nearly $70\%$ in computational wall time, while preserving deviations in $C_{\text{T}}$ and $C_{\text{Q}}$ within $1\%$.

Based on the present analysis, a temporal resolution corresponding to a $5^{\circ}$ azimuthal increment ($N_\text{steps} = 72$) combined with $\Delta x_\text{tip} = 2.5^{\circ}$ provides an effective balance between temporal accuracy, numerical robustness, and computational efficiency. Unless otherwise specified, this configuration is adopted for all subsequent validation cases.

\section{Case validation} \label{sec4}
Following the numerical convergence, robustness, and efficiency assessment presented in Sec.~\ref{sec3}, this section evaluates the predictive capability of the NL-UVLM-VPM solver with the adaptive conversion strategy under progressively more complex aerodynamic conditions. In contrast to previous studies \cite{Proulx-Cabana2022} in which validation was limited to hover conditions, the present assessment extends the validation envelope to forward flight and multi-rotor interaction scenarios.

Three benchmark rotor cases are considered. The first case, described in Sec.~\ref{sec4.1}, considers the two-bladed Caradonna-Tung rotor in hover \cite{Caradonna1981}, serving as a baseline evaluation of steady performance and near wake predictions. The second case, presented in Sec.~\ref{sec4.2}, investigates the full-scale AH-1G main rotor in forward flight \cite{Cross1988}, characterized by pronounced unsteady loading and BVI phenomena. Finally, Sec.~\ref{sec4.3} studies a side-by-side rotor configuration \cite{Sweet1960} to assess the solver's ability to capture rotor-rotor aerodynamic interference and induced flow interaction effects. Collectively, these cases examine the method’s robustness and predictive consistency across increasing levels of aerodynamic complexity.

\subsection{Hovering case: Caradonna-Tung rotor} \label{sec4.1}
The Caradonna and Tung (C-T) hovering rotor case \cite{Caradonna1981} is selected as a baseline validation configuration. The two-bladed rotor features untwisted rectangular blades with a NACA0012 airfoil and an aspect ratio of 6. The test conditions considered in this study are reported in Table.~\ref{table2}.
\captionsetup[table]{skip=1mm}
\renewcommand{\tabcolsep}{4mm}
\renewcommand{\arraystretch}{1.1}
\begin{table}[ht!]
\centering
\caption{Hover test conditions for C-T rotor\cite{Caradonna1981}}
\begin{tabular}{cc}
\hline
Collective pitch ($\mathrm{\theta_c}$) & Blade Tip Mach number ($\mathrm{M_{T}}$) \\
\hline
$5^{\circ}$ & \multirow{3}{*}{0.439} \\
$8^{\circ}$ & \\
$12^{\circ}$ & \\
\hline
\end{tabular}  \label{table2}
\end{table} 

High-fidelity URANS simulations are performed using the ROSITA solver \cite{Fu2022} on a Chimera grid system containing 8.65 million cells. A medium resolution grid, previously verified through grid-convergence studies, is adopted. Due to the time-accurate formulation, each revolution requires approximately 16 hours on 240 processors. In NL-UVLM-VPM simulations, a total of 20 rotor revolutions are computed to allow sufficient wake development and periodic convergence of the aerodynamic loads.

Fig.~\ref{fig7} compares the predicted integrated thrust coefficient $C_\text{T}$ with experimental measurements and URANS results for three collective pitch angles. The NL-UVLM-VPM predictions show close agreement with experimental data across all loading conditions and exhibit smaller deviations than the URANS results.
\begin{figure}[ht!]
\centering
\includegraphics[width=0.35\textwidth]{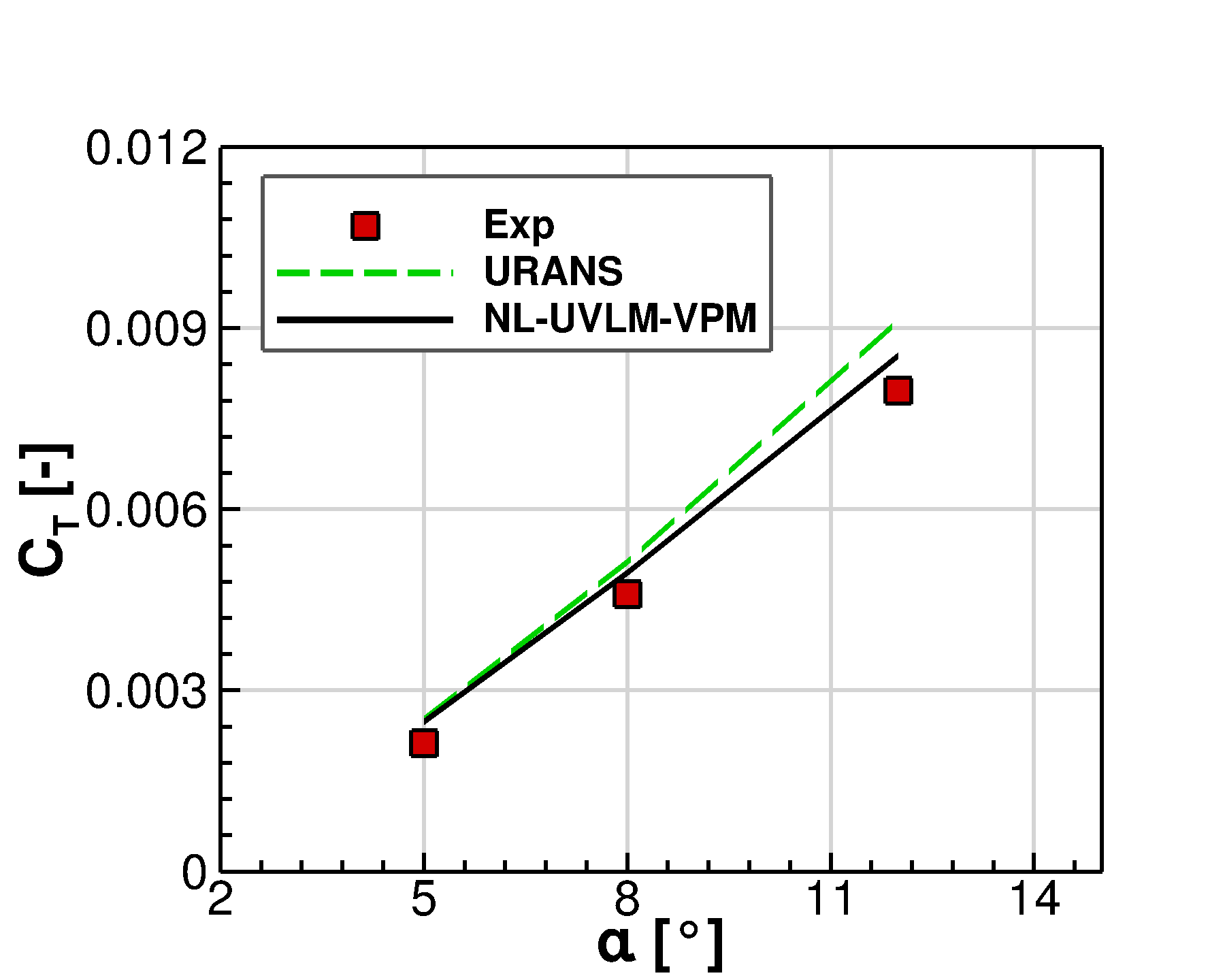}
\caption{Comparison of integrated rotor thrust coefficient on various collective pitch angles. \label{fig7}}
\end{figure} 

The chordwise pressure coefficient ($\mathrm{C_P}$) distributions at radial positions of $\mathrm{r/R} = 0.68$, $0.80$, and $0.89$ for $\mathrm{\theta_c = 5^{\circ}}$ are presented in Fig.~\ref{fig8}. In the NL-UVLM-VPM solution, pressure coefficients are reconstructed from the 2D RANS database using the effective angle of attack at each blade section. It can be seen that the predicted pressure distributions agree well with both experimental and URANS data for all radial positions. However, minor discrepancies become more significant toward the blade tip, primarily due to three-dimensional effects and strong tip-vortex influence.
\begin{figure*}[ht!]
\centering
\begin{subfigure}{0.35\textwidth}
\includegraphics[width=\linewidth]{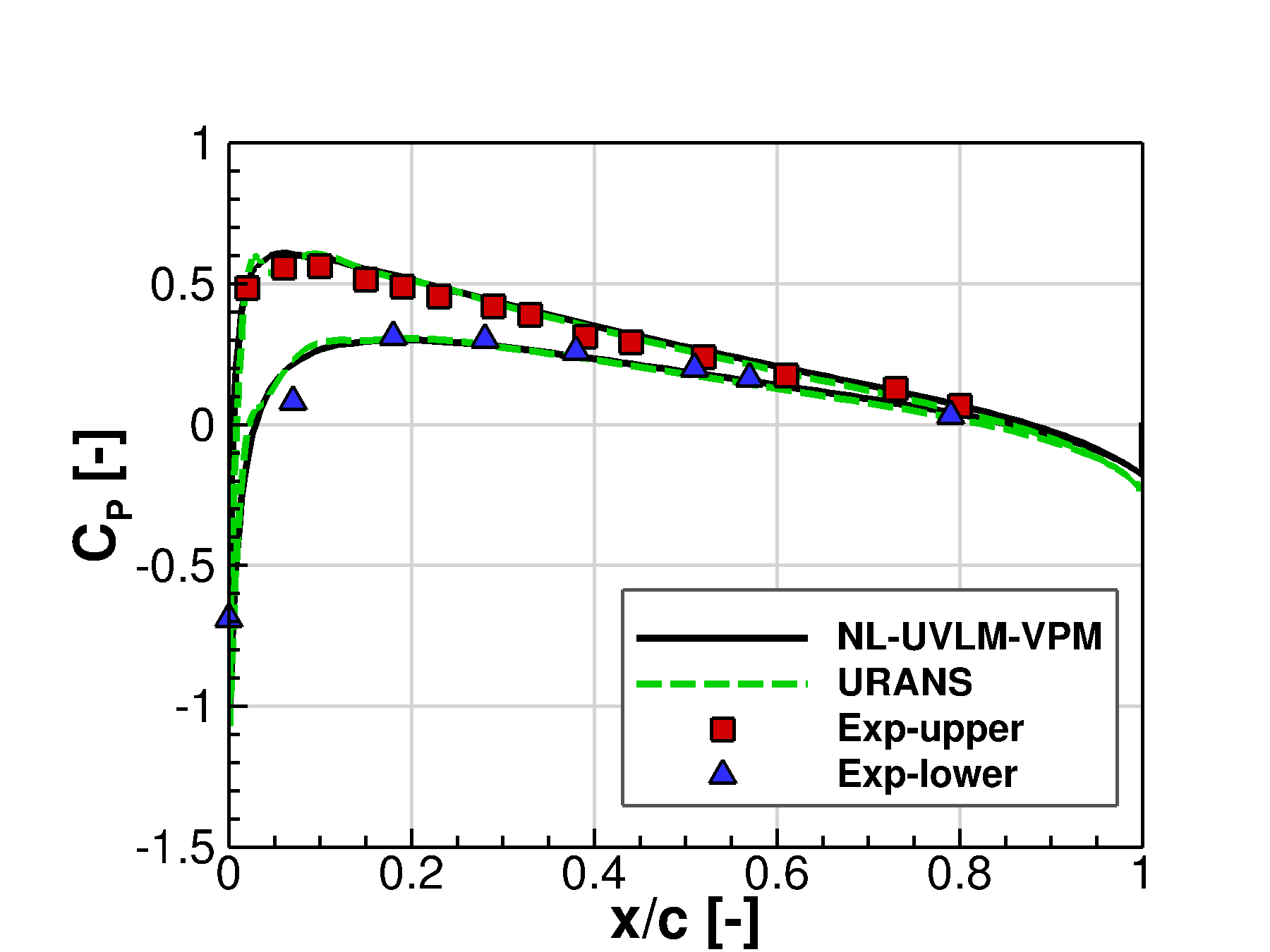}
\caption{r/R = 0.68} 
\end{subfigure}
\hspace{5.0mm}
\begin{subfigure}{0.35\textwidth}
\includegraphics[width=\linewidth]{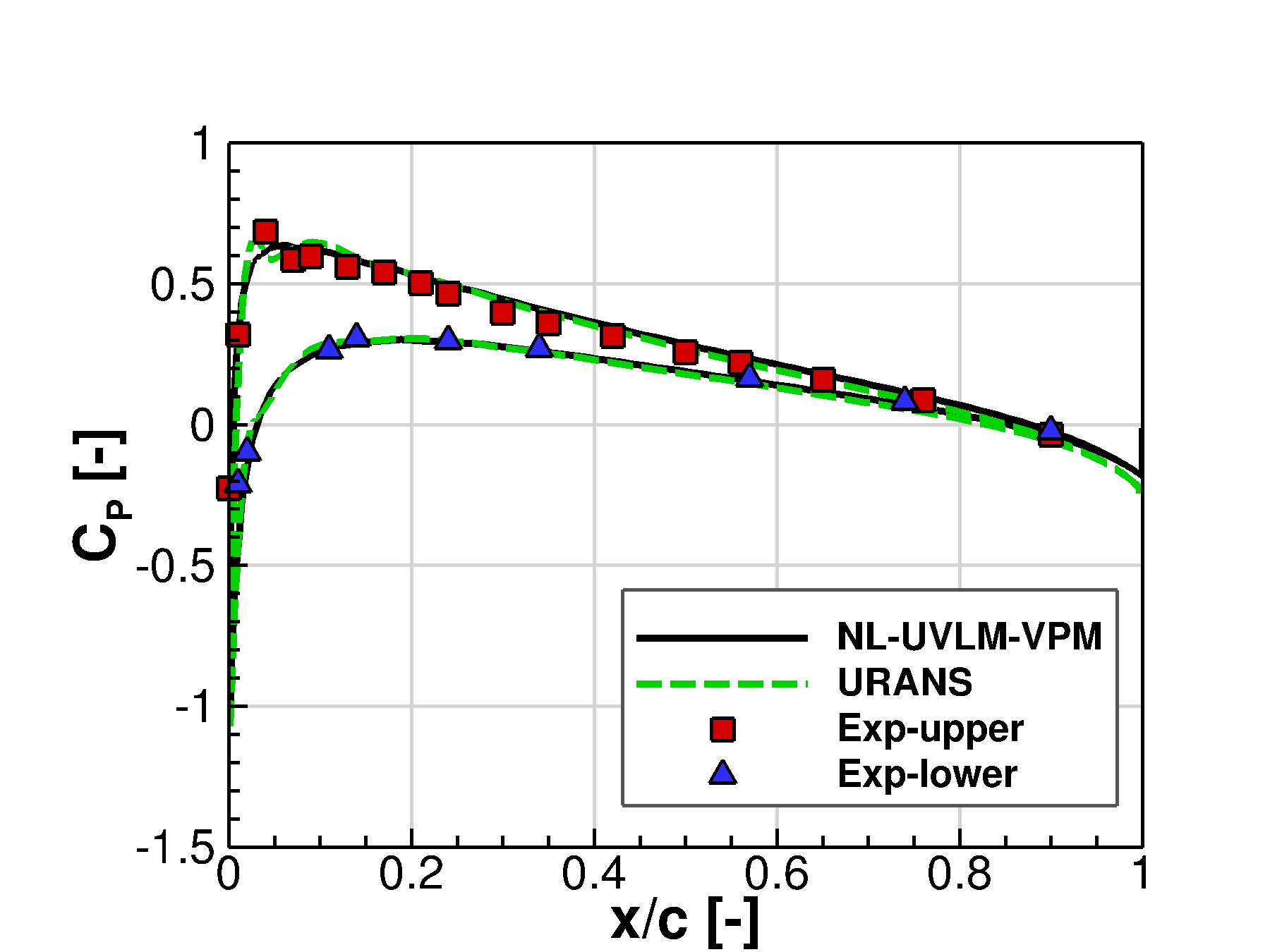}
\caption{r/R = 0.80}
\end{subfigure}
\hspace{5.0mm}
\begin{subfigure}{0.35\textwidth}
\includegraphics[width=\linewidth]{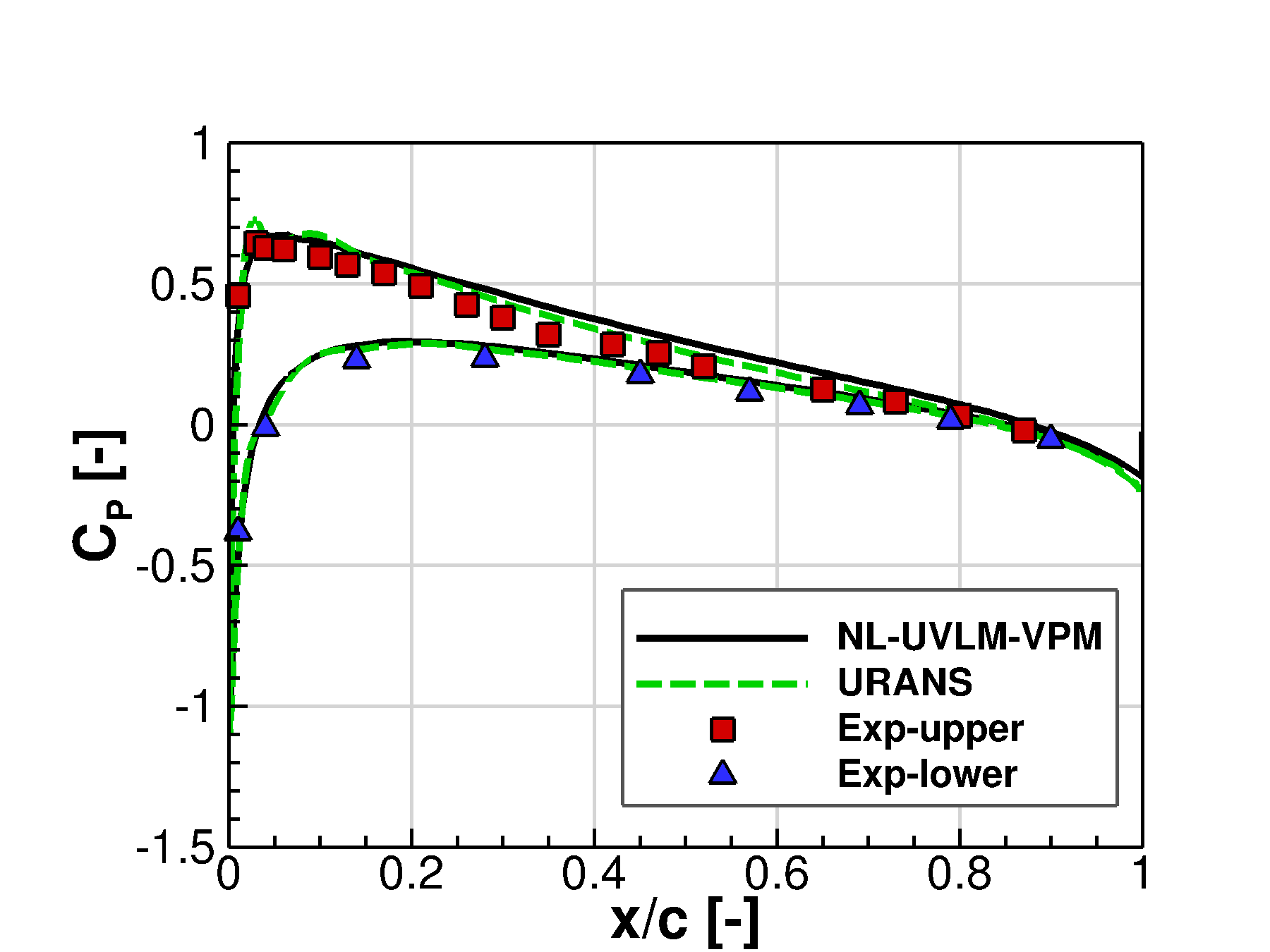}
\caption{r/R = 0.89}
\end{subfigure}
\hspace{5.0mm}
\begin{subfigure}{0.35\textwidth}
\includegraphics[width=\linewidth]{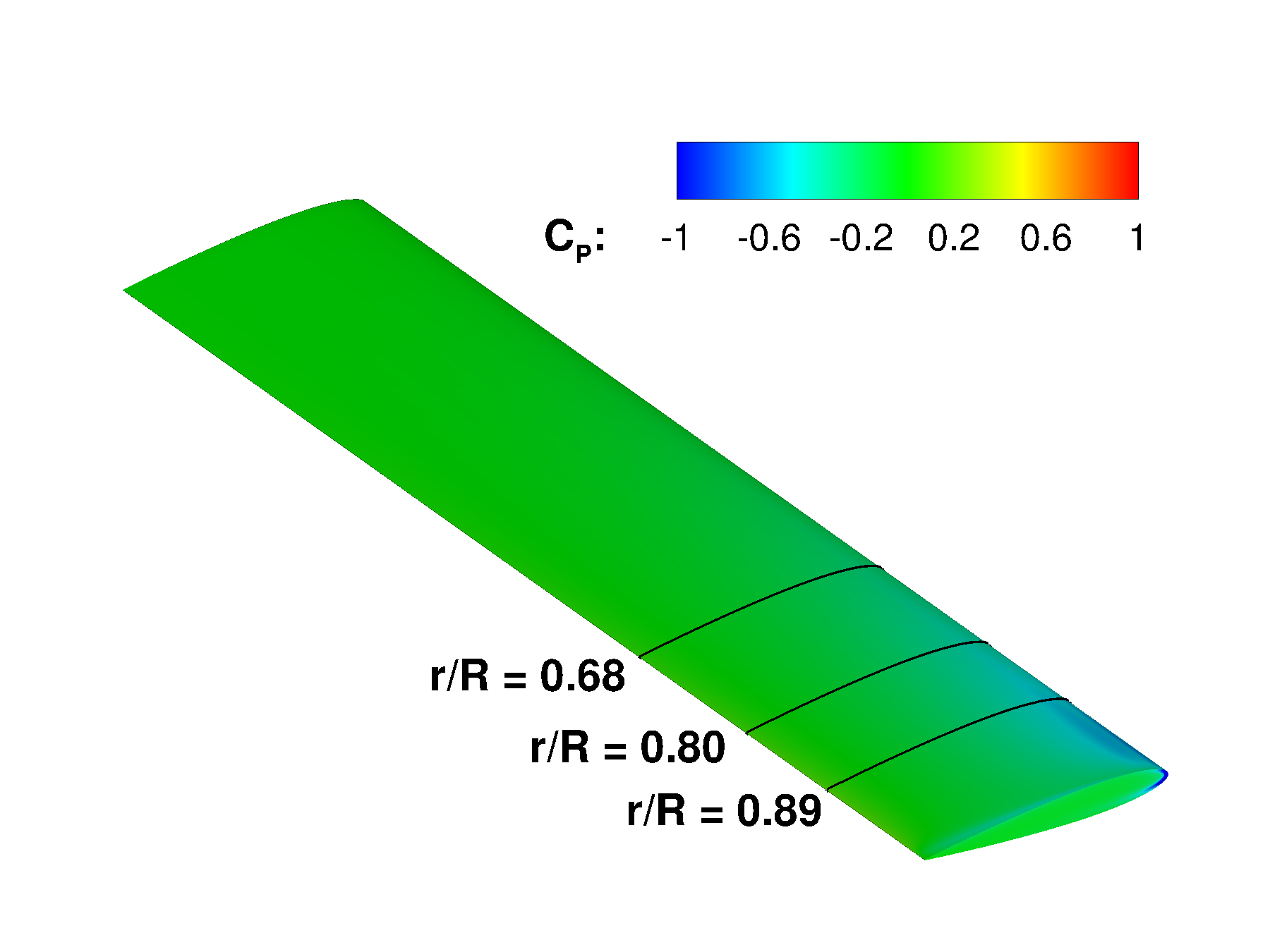}
\caption{Radial positions}
\end{subfigure}
\caption{$\mathrm{C_P}$ distribution over blade chord in different radial position for C-T rotor blade at $\mathrm{\theta_c = 5^{\circ}}$.\label{fig8} }
\end{figure*}

Fig.~\ref{fig9} compares the predicted tip vortex trajectories with experimental measurements and URANS results at three collective pitch settings. The NL-UVLM-VPM method captures the radial contraction and axial convection trends of the tip vortex across all loading conditions. At higher collective pitch, both numerical approaches present increasing deviation from experimental trajectories, reflecting the sensitivity of wake geometry to induced velocity modeling at higher thrust levels. Nevertheless, the overall contraction and descent trends remain in good agreement. These results indicate that the NL-UVLM-VPM solver with the adaptive conversion strategy reliably resolves rotor performance and near-wake geometry in hover flight.
\begin{figure*}[ht!]
\centering
\begin{subfigure}{0.35\textwidth}
\includegraphics[width=\linewidth]{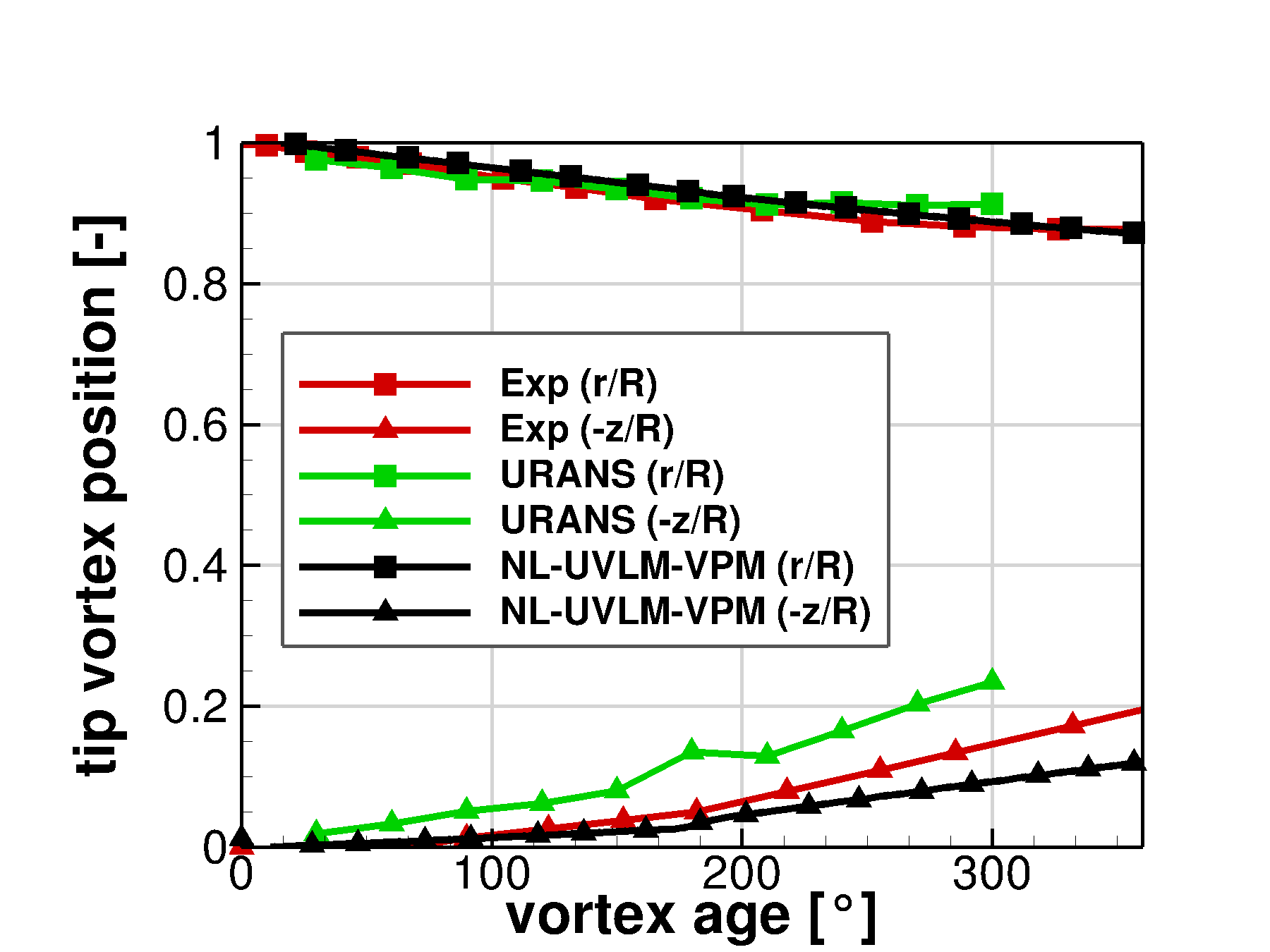}
\caption{$\mathrm{\theta_c} = 5^{\circ}$} 
\end{subfigure}
\hspace{5.0mm}
\begin{subfigure}{0.35\textwidth}
\includegraphics[width=\linewidth]{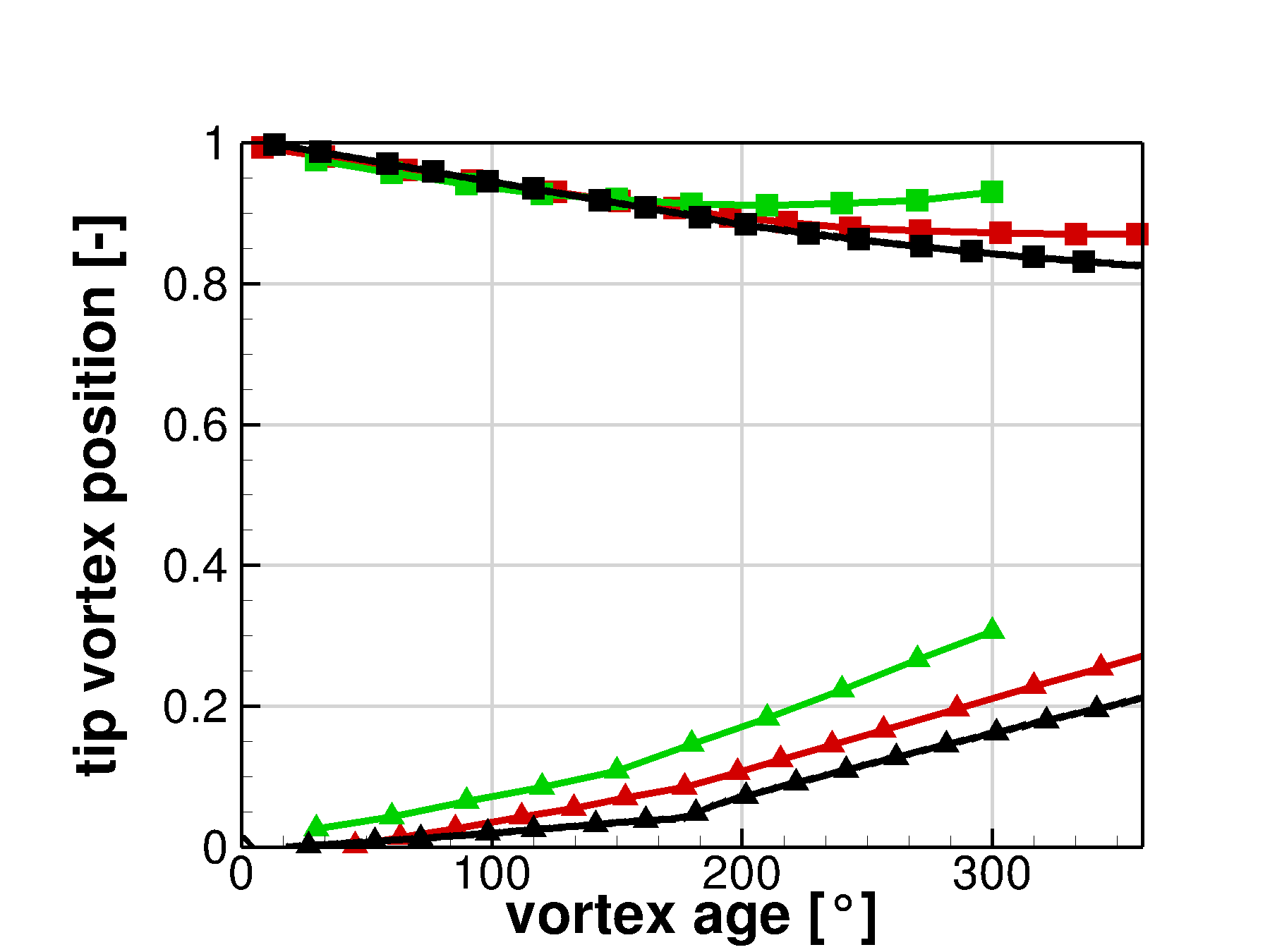}
\caption{$\mathrm{\theta_c} = 8^{\circ}$}
\end{subfigure}
\hspace{5.0mm}
\begin{subfigure}{0.35\textwidth}
\includegraphics[width=\linewidth]{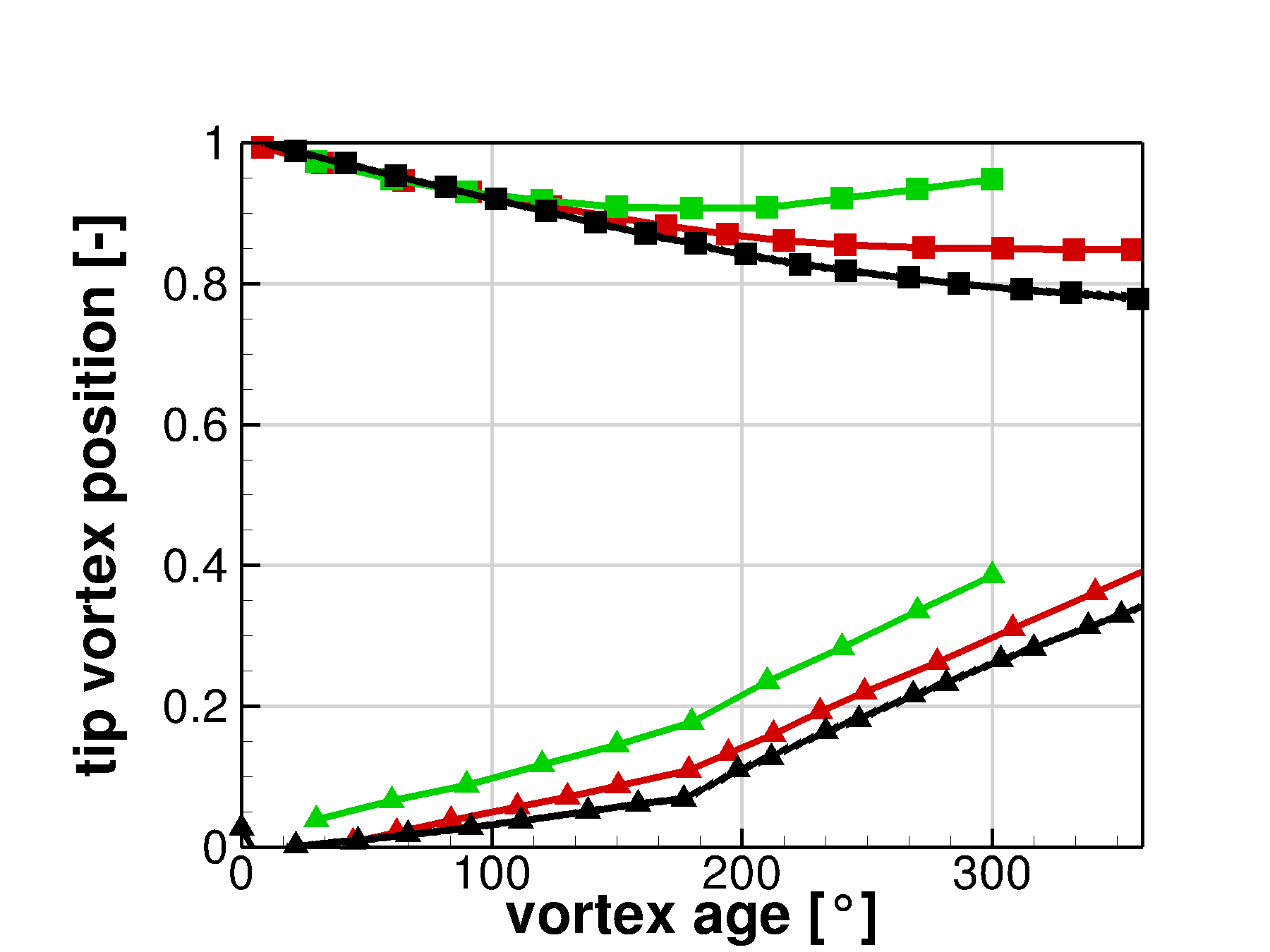}
\caption{$\mathrm{\theta_c} = 12^{\circ}$}
\end{subfigure}
\caption{Tip vortex position of C-T rotor for different collective pitch $\mathrm{\theta_c}$.\label{fig9} }
\end{figure*}

\subsection{Forward flight case: AH-1G main rotor} \label{sec4.2}
This section evaluates the predictive capability of the mid-fidelity framework under forward flight conditions. The AH-1G main rotor test case \cite{Cross1988} is considered as a representative configuration featuring pronounced unsteady aerodynamic loading and BVI phenomena.

The rotor geometry is summarized in Table.~\ref{table3}. The blade pitch and flapping motions are prescribed by the flight test measurements, as given in Eq.~\eqref{eq12}. The low-speed operating condition corresponds to test point 2157, with an advance ratio of 0.19 and a blade tip Mach number of 0.65.
\captionsetup[table]{skip=1mm}
\renewcommand{\tabcolsep}{4mm}
\renewcommand{\arraystretch}{1.1}
\begin{table}[ht!]
\centering
\caption{Geometric properties \cite{Cross1988}}
\begin{tabular}{lc}
\hline
Rotor property & Given value \\
\hline
$N_b$ & 2 \\
Blade planform & Rectangular \\
Blade section & OLS/TAAT\\
R & 6.7 m\\
c & 0.7283 m\\
$\theta_{tw}$ & Linear, $-10.0^\circ$\\
$\mathrm{R_c}$  & $\mathrm{0.154R}$ \\
\hline
\end{tabular}  \label{table3}
\end{table} 
\begin{equation} \label{eq12}
 \begin{split} 
    \theta(\psi) &= 6.0 + 1.7\cos(\psi) - 5.5\sin(\psi) \\
    \beta(\psi) &= 2.13\cos(\psi) - 0.15\sin(\psi)
 \end{split}
\end{equation}

High-fidelity URANS solutions are obtained using the ROSITA URANS solver \cite{Fu2023.1} on a Chimera grid system comprising 11.1 million cells. Each revolution requires approximately 10 hours on 360 cores. In the mid-fidelity analysis, the 2D viscous database was obtained from the CHAMPS solver \cite{Parenteau2020}, covering an appropriate range of Reynolds numbers, Mach numbers, and angles of attack.

Fig.~\ref{fig10} presents the time history of the integrated thrust coefficient over ten revolutions. After approximately two revolutions, both URANS and NL-UVLM-VPM solutions reach a quasi-periodic state with nearly identical amplitudes and phase variations. The two predictions remain closely overlapped throughout subsequent revolutions, indicating that the present mid-fidelity method accurately reproduces the unsteady thrust response observed in the URANS simulation under forward-flight conditions.
\begin{figure}[ht!]
\centering
\includegraphics[width=0.8\textwidth]{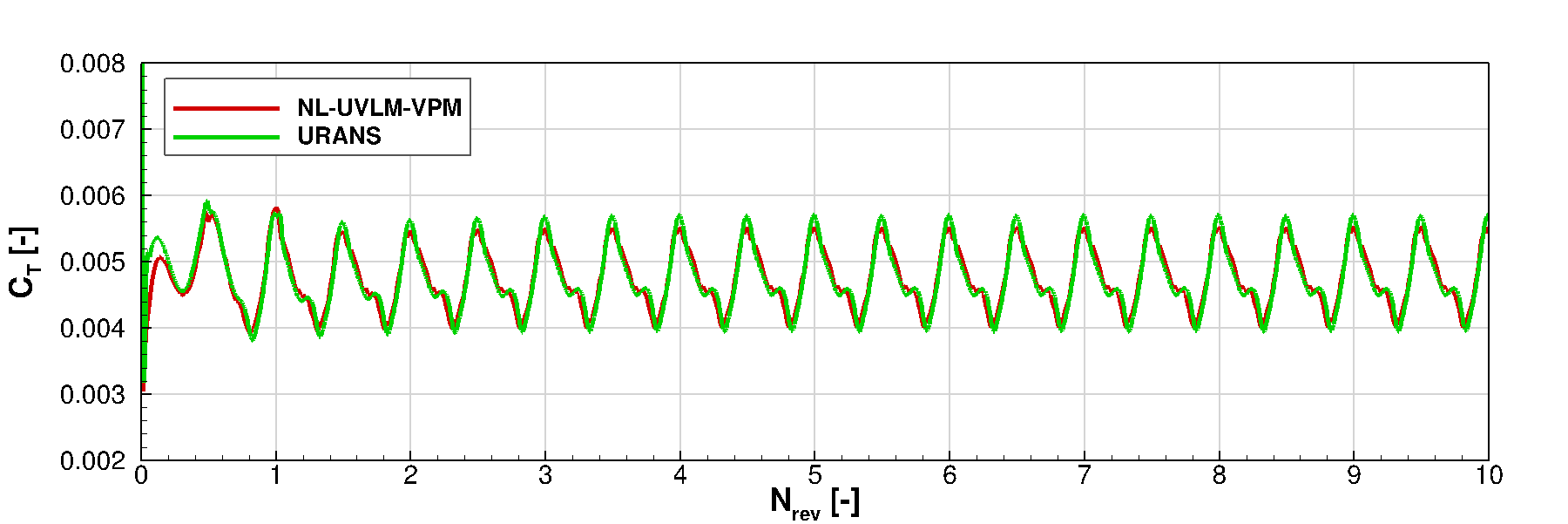}
\caption{Time history of the integrated thrust coefficient $\mathrm{C_T}$ over 10 rotor revolutions for the AH-1G rotor in forward flight. \label{fig10}}
\end{figure} 

For comparison with the flight test measurements, the thrust coefficients were smoothed using a simple moving average (SMA) method. This post-processing procedure eliminates oscillations and enables a direct comparison of the average aerodynamic trends across three datasets. As summarized in Table.~\ref{table4}, both numerical results slightly overpredict the experimental value, with deviations below $3\%$. The NL-UVLM-VPM prediction closely correlated with the URANS result, confirming its reliability for overall thrust performance prediction in forward flight.
\captionsetup[table]{skip=1mm}
\renewcommand{\tabcolsep}{4mm}
\renewcommand{\arraystretch}{1.1}
\begin{table}[ht!]
\centering
\caption{Integrated thrust coefficient comparisons}
\begin{tabular}{lcc}
\hline
Method & $C_T$ & $|$ Errors $|$ \\
\hline
Experiment\cite{Cross1988} & 0.00464 & - \\
URANS & 0.00476 & 2.6$\%$ \\
NL-UVLM-VPM & 0.00478 & 3.0$\%$ \\ 
\hline
\end{tabular} \label{table4}
\end{table} 

To assess the accuracy of the mid-fidelity simulation at a local level, Fig.~\ref{fig11} shows the chordwise pressure coefficient distributions at several azimuthal positions ($\mathrm{\psi} = 75^{\circ}$, $90^{\circ}$, $105^{\circ}$, $270^{\circ}$, $285^{\circ}$, and $300^{\circ}$) for the specific radial station ($\mathrm{r/R = 0.91}$). Overall, the NL-UVLM-VPM predictions agree well with both flight measurements and URANS results across the advancing and retreating sides. On the advancing side, both numerical methods capture the suction peak near the leading edge, which is associated with the high dynamic pressure as the advancing blade experiences increased relative inflow velocity. On the retreating side, a strong suction spike is observed experimentally, which is reproduced more clearly by URANS, while the NL-UVLM-VPM solution shows a slightly attenuated peak. Nevertheless, both approaches follow the overall pressure recovery trend.
\begin{figure}[ht!]
\centering
\includegraphics[width=0.95\textwidth]{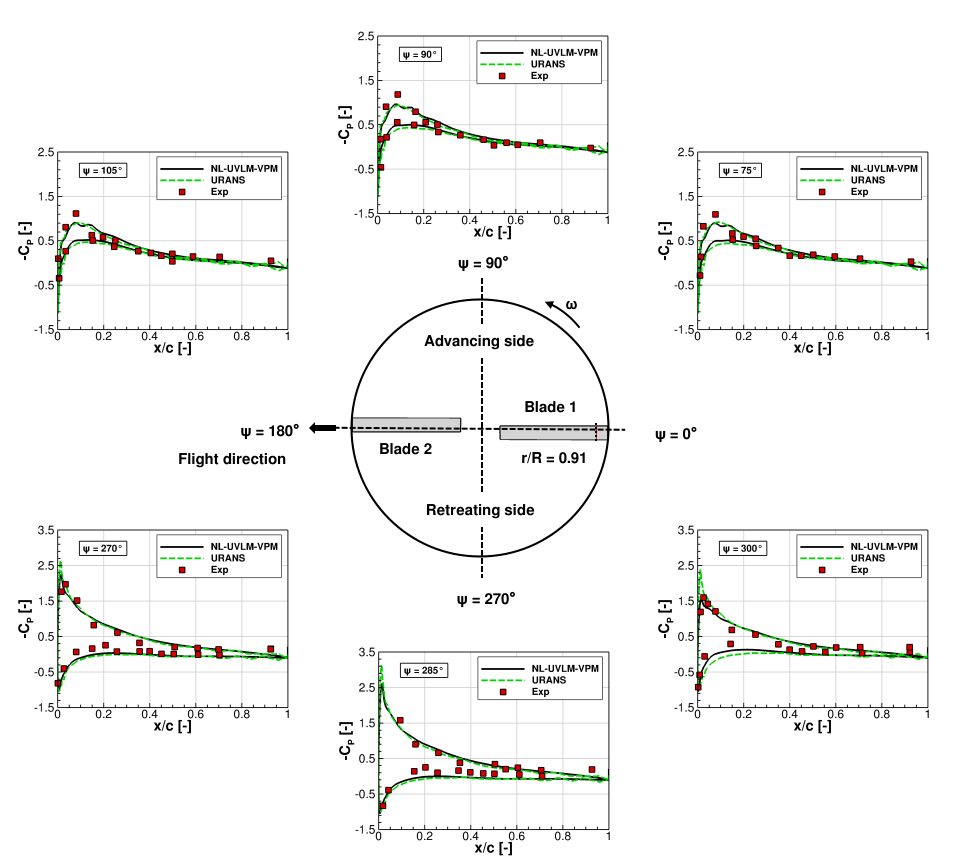}
\caption{Pressure coefficient distribution at r/R = 0.91 for AH-1G main rotor in forward flight. \label{fig11}}
\end{figure} 

Fig.~\ref{fig12} illustrates the sectional thrust coefficient variation at r/R = 0.75 and 0.91 over one rotor revolution. The NL-UVLM-VPM predictions show good agreement with both the flight test measurements and URANS simulations in terms of magnitude and phase across most azimuthal angles. Both numerical approaches overpredict the thrust on the advancing ($0^{\circ}$ to $90^{\circ}$) and retreating ($300^{\circ}$ to $360^{\circ}$) sides. These discrepancies are likely related to the absence of aeroelastic effects on the rotor blades in the present simulations, which can alter the effective angle of attack and the phase of the blade loads during flight. BVI signatures are observed near azimuthal angles of approximately $60^{\circ}$ to $120^{\circ}$ and around $300^{\circ}$. The NL-UVLM-VPM solution captures these fluctuations, although their amplitude is somewhat underpredicted. The URANS results show smoother behavior in these regions, consistent with the numerical dissipation inherent to grid-based methods.
\begin{figure*}[ht!]
\centering
\begin{subfigure}{0.35\textwidth}
\includegraphics[width=\linewidth]{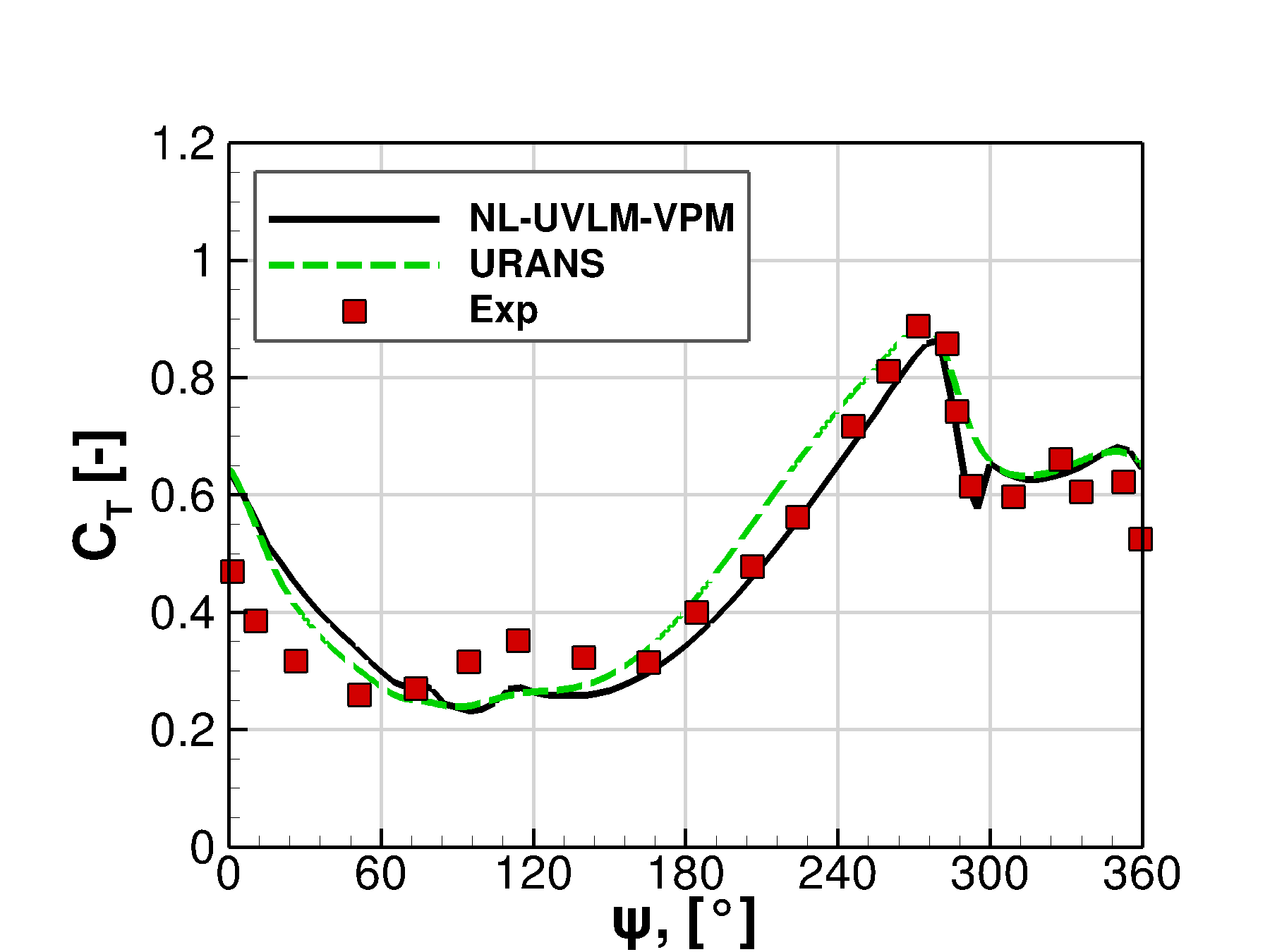}
\caption{$\mathrm{r/R=0.75}$} 
\end{subfigure}
\hspace{5.0mm}
\begin{subfigure}{0.35\textwidth}
\includegraphics[width=\linewidth]{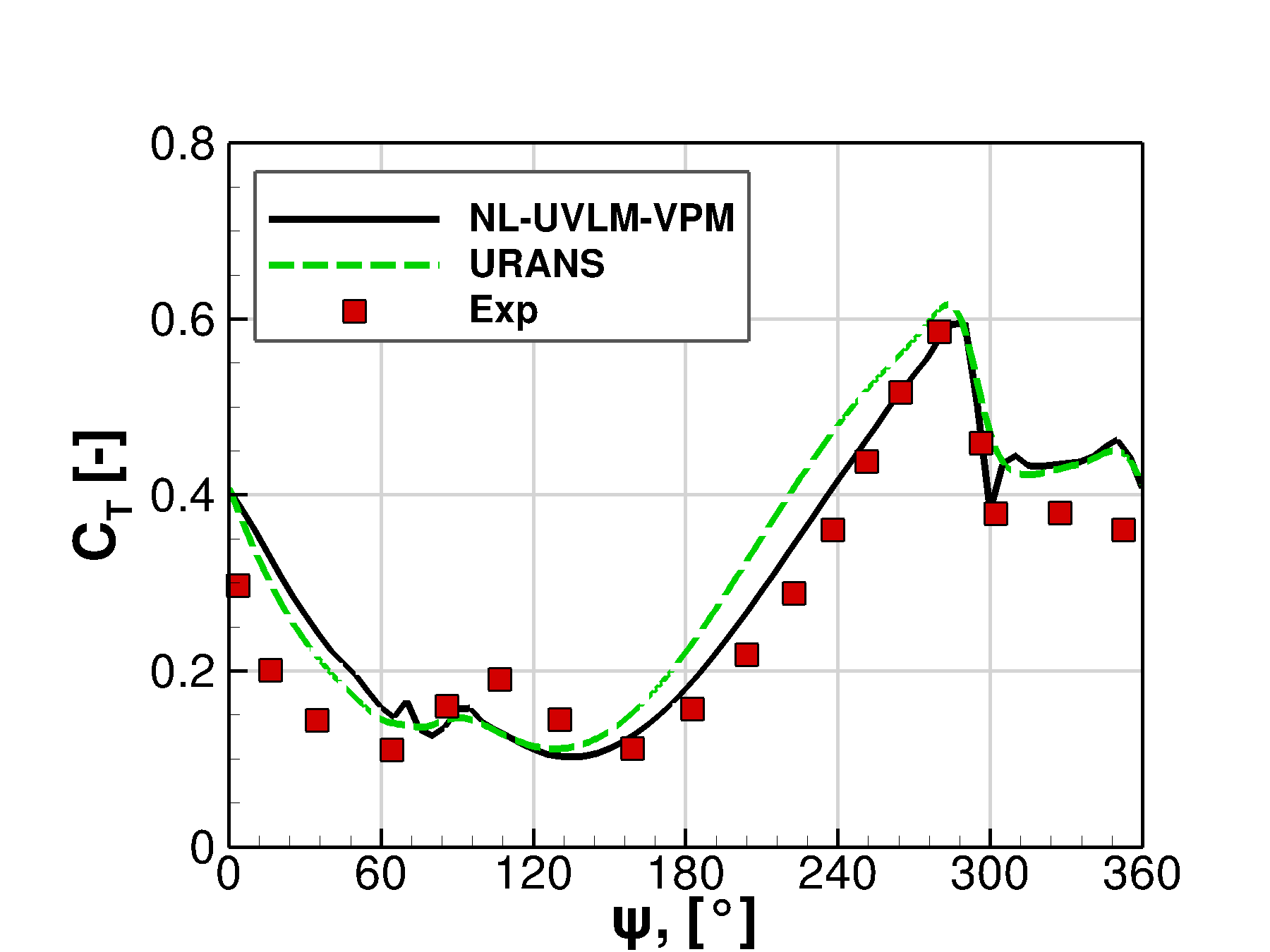}
\caption{$\mathrm{r/R=0.91}$}
\end{subfigure}
\caption{Blade sectional thrust coefficients history over one rotor revolution. \label{fig12} }
\end{figure*}

The global wake topology is illustrated in Fig.~\ref{fig13}, where vortex particles are colored by turbulent viscosity $\mathrm{\nu_t}$. It can be seen that the roll-up of the tip vortices on both the advancing and retreating sides, as well as their downstream convection, is clearly reproduced. The wake deformation in the forward flight region and the interaction between successive blade-tip vortices are also captured, demonstrating that the Lagrangian wake treatment maintains coherent vortex structures in forward flight.
\begin{figure}[ht!]
\centering
\includegraphics[width=0.5\textwidth]{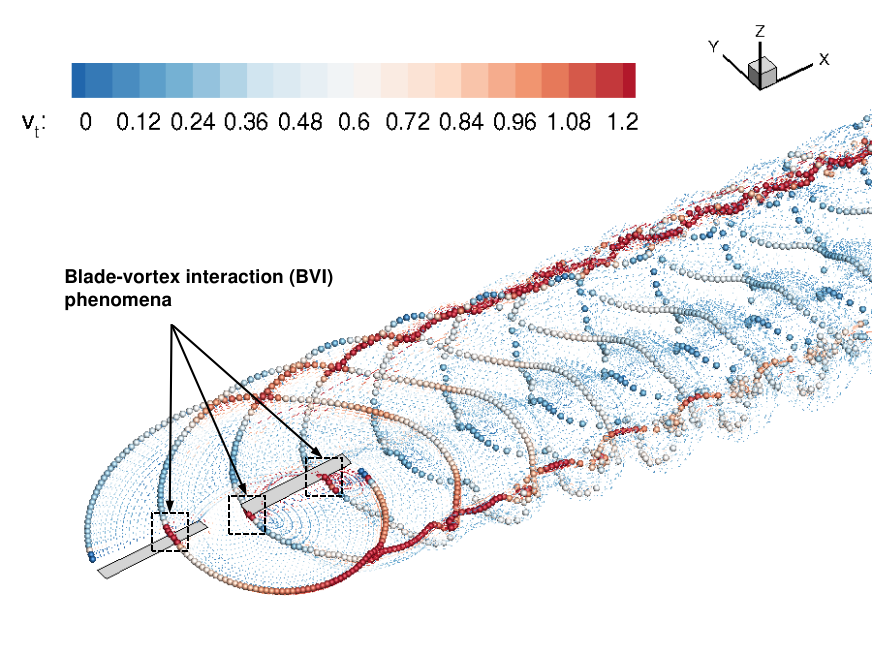}
\caption{Wake structures of AH-1G main rotor in forward flight. \label{fig13}}
\end{figure} 

\subsection{Rotor-rotor interaction case: Large-scale side-by-side rotor} \label{sec4.3}
Building upon the single-rotor validations, this section assesses the mid-fidelity approach for multi-rotor aerodynamic interaction. Experimental data from Sweet's side-by-side rotor test in hover \cite{Sweet1960} are used for validation. The configuration consists of two identical counter-rotating rotors with an overlap ratio L/R = 2.03 and a phase difference of 90$^\circ$, as shown in Fig.~\ref{fig14}. The present study focuses on the high-solidity rotor configuration. The geometric and operating parameters are summarized in Table.~\ref{table5}.
\begin{figure}[ht!]
\centering
\includegraphics[width=0.45\textwidth]{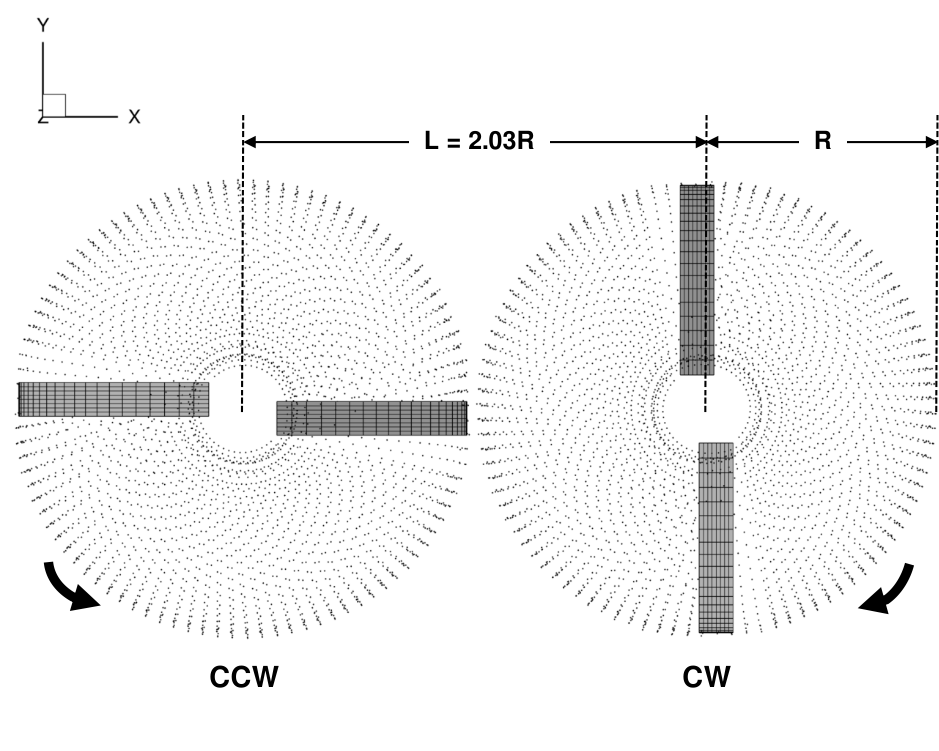}
\caption{Side-by-side rotor configuration, L/R = 2.03. \label{fig14}}
\end{figure} 

\captionsetup[table]{skip=1mm}
\renewcommand{\tabcolsep}{4mm}
\renewcommand{\arraystretch}{1.1}
\begin{table}[ht!]
\centering
\caption{Geometric properties and operating parameters\cite{Sweet1960}}
\begin{tabular}{lcc}
\hline
Rotor property & Left rotor & Right rotor \\
\hline
$N_b$ & 2 & 2\\
Blade planform & Rectangular & Rectangular\\
Blade section & NACA0012 & NACA0012\\
R & 15.25 feet & 15.25 feet\\
c & 1.16 feet & 1.16 feet\\
$\theta_{tw}$ & 0$^\circ$ & 0$^\circ$\\
$R{_c}$  & 0.151R & 0.151R \\
Rotational speed & 626.18 RPM & -626.18 RPM \\
\hline
\end{tabular}  \label{table5}
\end{table} 

High-fidelity reference solutions are obtained using time-accurate URANS simulations on a Chimera grid system comprising approximately 17.3 million cells. Each simulation requires three days to complete five rotor revolutions on 640 cores. Mid-fidelity simulations are performed for both an isolated single-rotor baseline and the side-by-side configuration. Hover conditions with varying collective pitch are considered to evaluate performance across different loading levels.

Fig.~\ref{fig15} compares the thrust-torque relationship and the figure of merit (FM) for the side-by-side configuration. The NL-UVLM-VPM predictions show good agreement with both experimental measurements and URANS results across the tested thrust range. At high loading conditions (large C${_T}$), where URANS data are available, both numerical approaches slightly underpredict $\mathrm{FM}$ relative to the experiment. This discrepancy may stem from the absence of aeroelastic deformation in the present simulations, which can affect rotor performance under elevated thrust levels. 
\begin{figure*}[ht!]
\centering
\begin{subfigure}{0.35\textwidth}
\includegraphics[width=\linewidth]{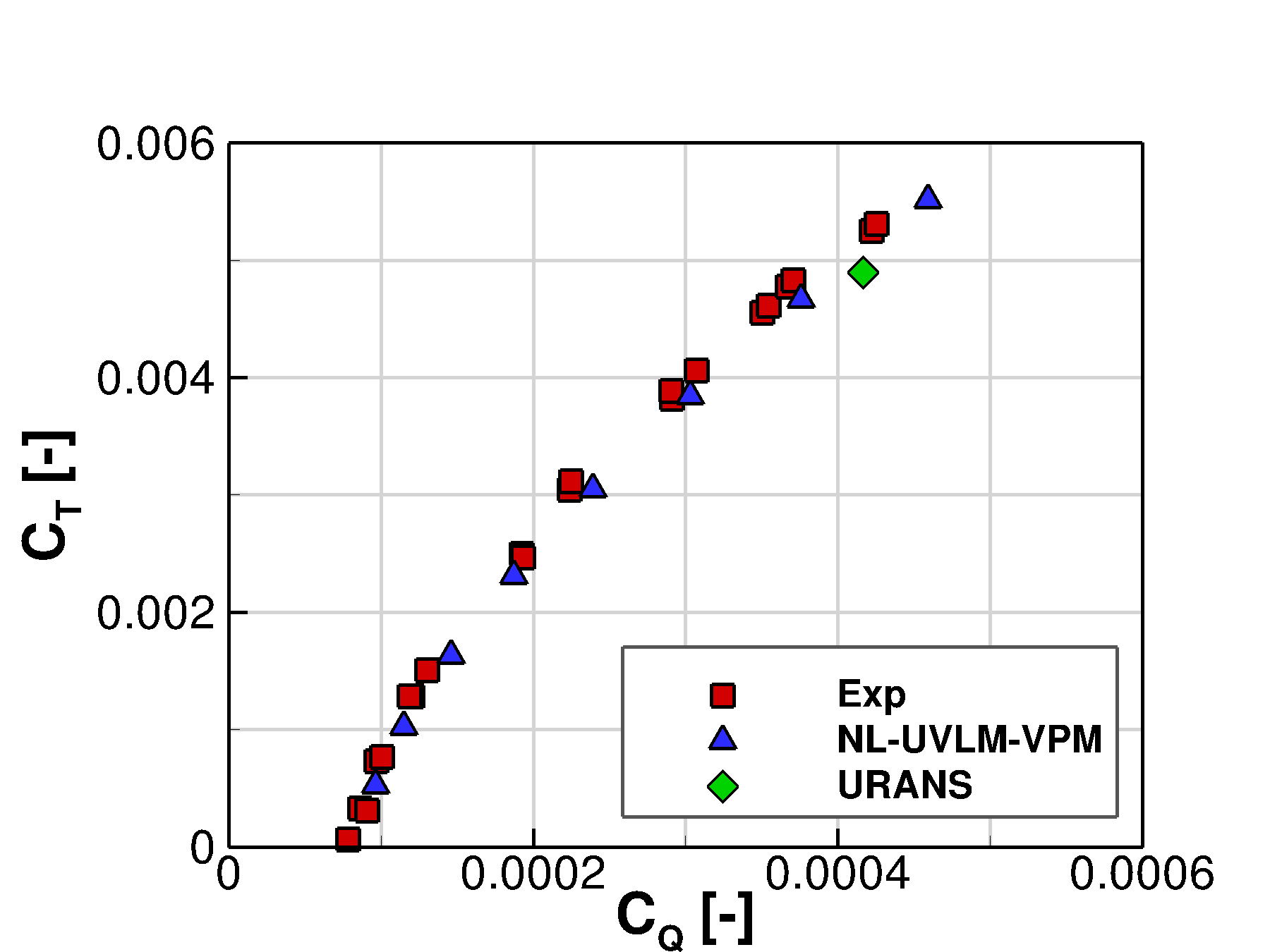}
\caption{$\mathrm{C_T}$ as a function of $\mathrm{C_Q}$} 
\end{subfigure}
\hspace{5.0mm}
\begin{subfigure}{0.35\textwidth}
\includegraphics[width=\linewidth]{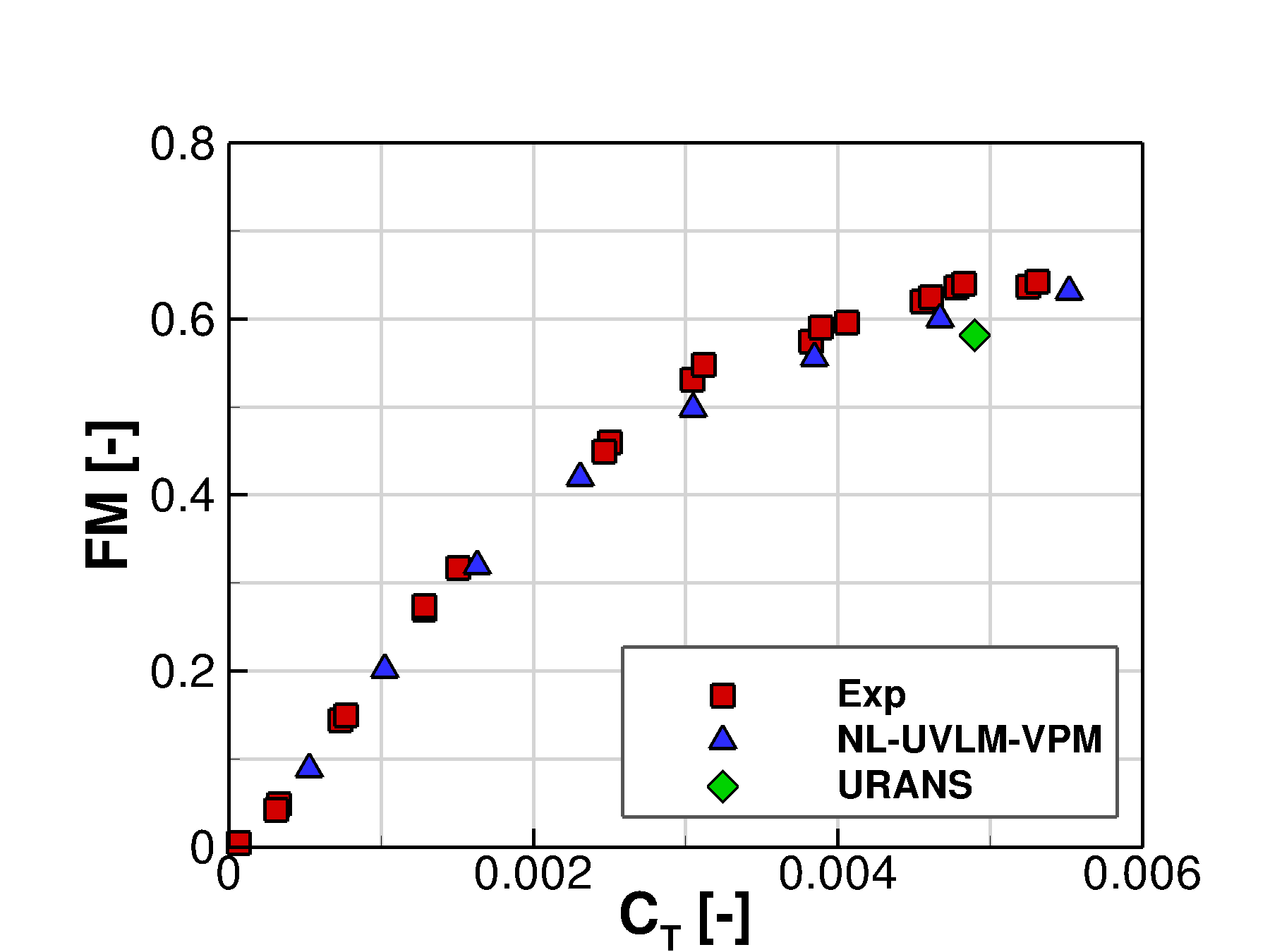}
\caption{Figure of merit FM as a function of $\mathrm{C_T}$}
\end{subfigure}
\caption{Comparison of rotor performance for the side-by-side rotor model (L/R = 2.03). \label{fig15} }
\end{figure*}

Fig.~\ref{fig16} presents the spanwise distribution of the normal force coefficient (C$_N$) at three azimuthal phases for $\theta_c= 8^\circ$, with the single-rotor solution used as a reference. These results provide insight into the unsteady evolution of blade loading as the rotor interacts with the skewed wake and upwash field induced by the adjacent rotor. On the advancing side ($0^{\circ} \le \psi \le 90^{\circ}$), blade loading gradually recovers as the blade leaves the interaction region. During the transition phase ($90^{\circ} \le \psi \le 270^{\circ}$), the loading remains close to the single-rotor reference. Upon re-entry into the interference region on the retreating side ($270^{\circ} \le \psi \le 360^{\circ}$), noticeable inboard loading increase and outboard loading reduction are observed, reflecting the asymmetric induced flow field generated by rotor mutual interaction.
\begin{figure}[ht!]
\centering
\includegraphics[width=0.45\textwidth]{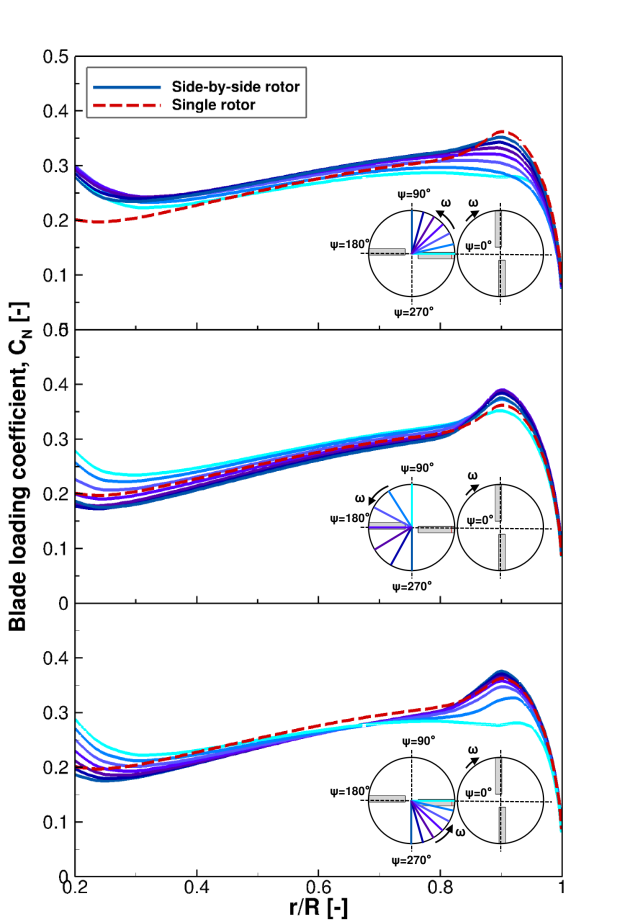}
\caption{Blade loading as blade goes through region of wake skewness and accentuated upwash between the rotors. \label{fig16}}
\end{figure} 

The colormaps of the instantaneous axial velocity fields at N$_{\text{rev}}$ = 10, together with the velocity profiles extracted at z/R = -0.5 and z/R = -1.0, are displayed in Fig.~\ref{fig17}, where the URANS results are included for comparison. The contours clearly capture two dominant downwash regions beneath the rotors and a weaker interaction region between them. As the wake convects downstream, the induced velocity cores move inward and partially merge, indicating wake contraction and mutual induction. This interaction produces a mildly asymmetric velocity distribution with localized upwash in the midplane, as reflected in the extracted velocity profiles. Meanwhile, the quantitative comparison of velocity profiles indicates a consistent representation of the dominant interaction-induced flow features between NL-UVLM-VPM and URANS.
\begin{figure}[ht!]
\centering
\includegraphics[width=0.8\textwidth]{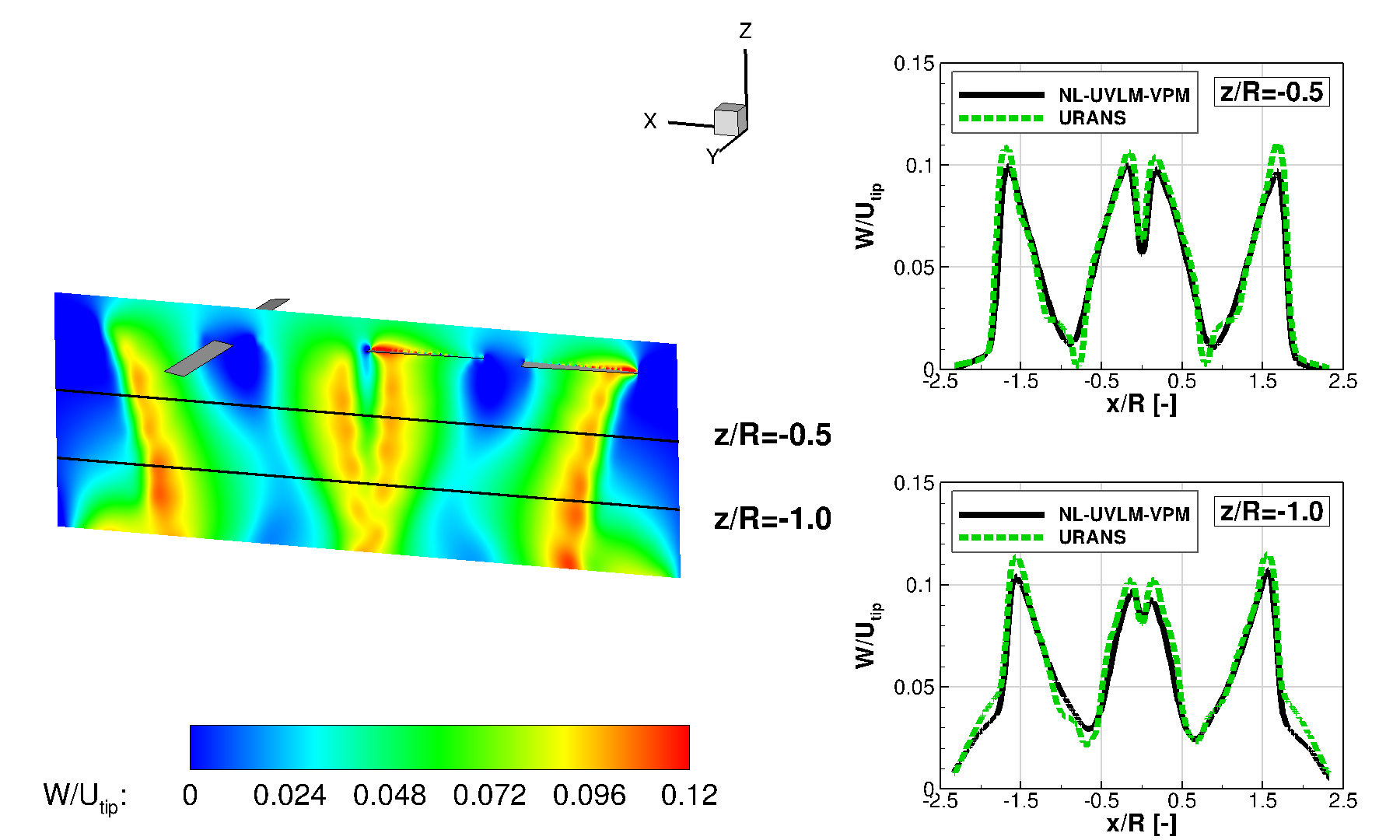}
\caption{Contour of the instantaneous velocity along the z-axis and velocity profiles
extracted at z/R = -0.5 and -1.0. \label{fig17}}
\end{figure} 

Fig.~\ref{fig18} shows the instantaneous vortical structures identified using the Q-criterion (Q = 0.005) for both approaches. It can be seen that both simulations reproduce similar coherent helical tip vortices that roll up and convect downstream. A distinct inward contraction of the wake is observed, confirming the presence of mutual rotor induction. The NL-UVLM-VPM solution preserves dominant wake behavior, including tip-vortex roll-up, rotor mutual interaction, and wake skewness, in a manner consistent with those predicted by the URANS simulation. These alignments demonstrate that the NL-UVLM-VPM approach effectively captures the essential physics of multi-rotor aerodynamic interaction.

\begin{figure*}[ht!]
\centering
\begin{subfigure}{0.4\textwidth}
\includegraphics[width=\linewidth]{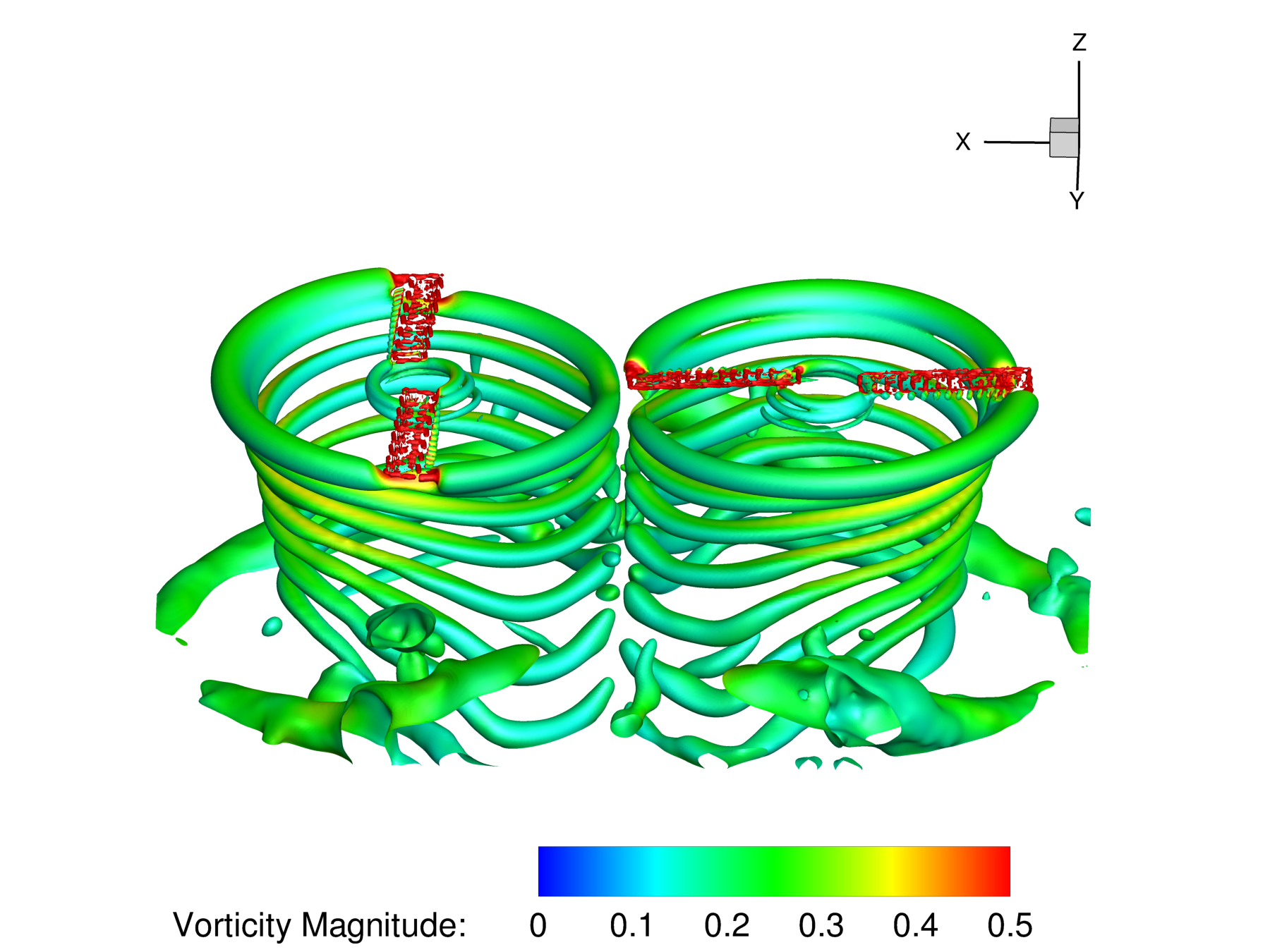}
\caption{NL-UVLM-VPM solution \label{fig20a}} 
\end{subfigure}
\hspace{5.0mm}
\begin{subfigure}{0.4\textwidth}
\includegraphics[width=\linewidth]{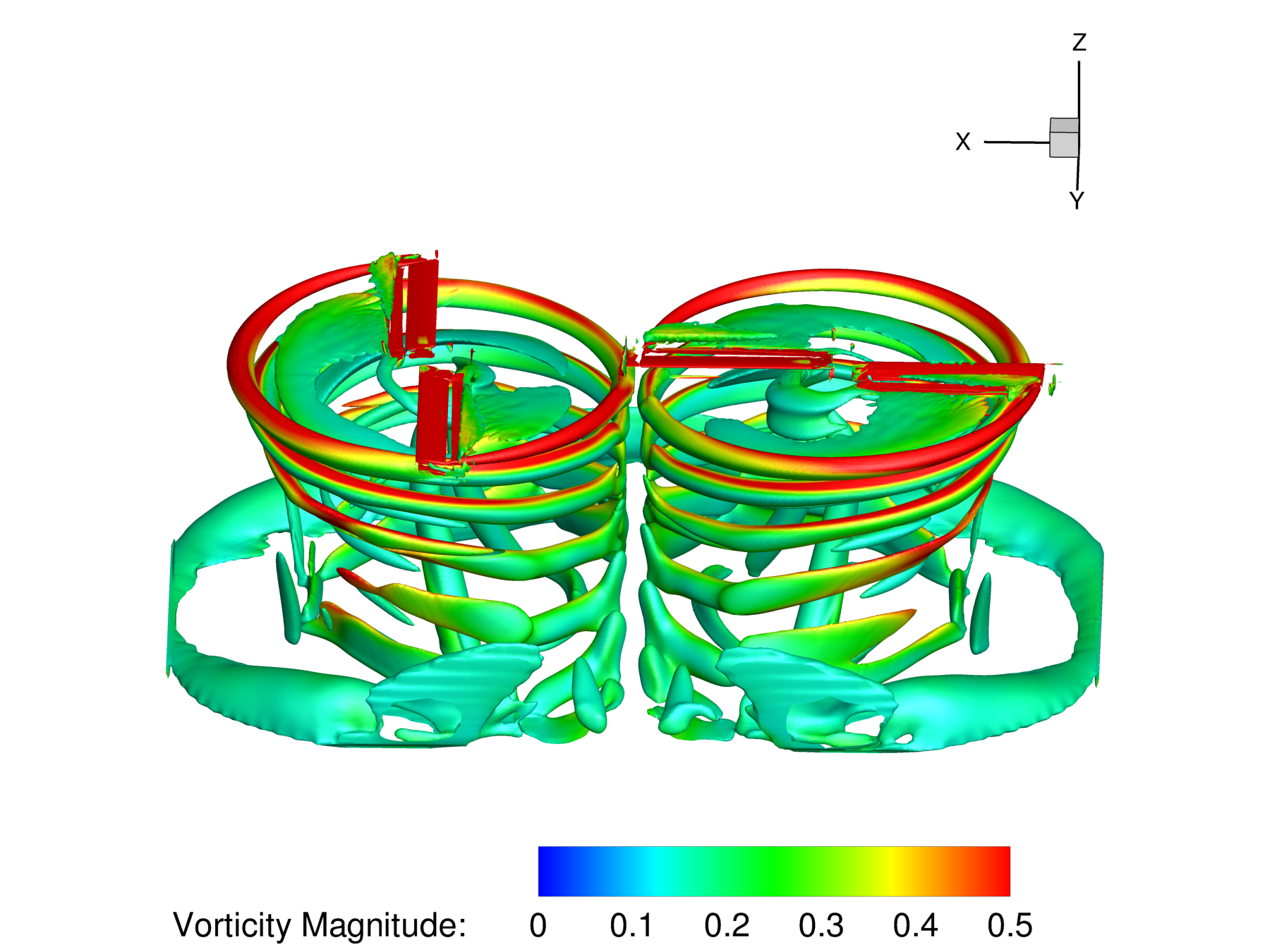}
\caption{URANS solution \label{fig20b}}
\end{subfigure}
\caption{Comparison between vortical structures obtained from NL-UVLM-VPM and URANS calculations.\label{fig18} }
\end{figure*}

\subsection{Computational time}
To assess computational efficiency, the average CPU time per revolution for the NL-UVLM-VPM and URANS simulations across all three benchmark cases is summarized in Table.~\ref{table6}. All computations were performed on the Niagara high-performance cluster at Compute Canada, which consists of 2024 nodes, each equipped with 40 Intel "Skylake" cores at 2.4 GHz, interconnected via an EDR Infiniband network.

The NL-UVLM-VPM approach requires significantly fewer computational resources than the corresponding time-accurate URANS simulations. Depending on the test case, the mid-fidelity solver achieves a speedup ratio exceeding two orders of magnitude. These results highlight the substantial reduction in computational cost offered by the mid-fidelity approach, while maintaining comparable predictions of aerodynamic loads and near-wake flow structures.
\captionsetup[table]{skip=1mm}
\renewcommand{\tabcolsep}{2.0mm}
\renewcommand{\arraystretch}{1.1}
\begin{table}[ht!]
\centering
\caption{CPU run-time of two approaches for all three test cases simulations}
\begin{tabular}{lcccc}
\hline
Test case & Approach & CPU cores & CPUh/rev & Speedup\\
\hline
\multirow{2}{*}{C-T rotor} & NL-UVLM-VPM & 40 & 26 & $147.2$\\
& URANS & 240 & 3827 & $1.0$\\
\multirow{2}{*}{AH-1G rotor} & NL-UVLM-VPM & 40 & 25 & $142.4$\\
& URANS & 360 & 3560 & $1.0$\\
\multirow{2}{*}{Side-by-side rotor} & NL-UVLM-VPM & 40 & 53 & $173.9$\\
& URANS & 640 & 9216 & $1.0$\\
\hline
\end{tabular}  \label{table6}
\end{table} 

\section{Conclusions} \label{sec5}
The present study introduces a hybrid nonlinear unsteady vortex-lattice vortex-particle method with a scale-consistent adaptive wake panel-particle conversion strategy for rotorcraft aerodynamic simulations. The proposed strategy alleviates the inherent temporal–spatial resolution coupling of conventional wake treatments, enabling flexible control of wake spatial resolution while preserving the temporal accuracy of the underlying time-integration scheme. Numerical assessments showed that the adaptive conversion strategy maintains near third-order temporal convergence and improves numerical robustness under coarsened temporal resolution.

For representative hover simulations, this approach reduced computational wall-clock time by approximately $29\%$ relative to the conventional conversion strategy at identical temporal resolution, and by nearly 70$\%$ compared with a fine-resolution reference simulation over 20 rotor revolutions, while maintaining thrust and torque predictions within 1$\%$ of the reference solution. The study further provides practical recommendations for particle conversion parameters and wake resolution for numerical stability and computational efficiency.

The methodology was validated across a broadened range of operating conditions, including hover, forward flight with pronounced blade-vortex interaction, and side-by-side rotor aerodynamic interference. In these cases, good agreement with experimental measurements and dedicated high-fidelity URANS simulations was observed in integrated loads, sectional distributions, and dominant wake structures. In terms of computational cost, the present NL-UVLM-VPM framework achieves speedups exceeding two orders of magnitude in CPU-hours per revolution relative to time-accurate URANS for the cases considered, demonstrating the capability of this approach for accurate and efficient simulation of complex rotorcraft aerodynamic environments.

Future work will focus on extending the present method for complex tip geometries using the 2.5D hypothesis \cite{Lavoie2018}, incorporating rotor-fuselage interaction through panel method coupling, and integrating multidisciplinary effects such as aeroelasticity and aero-icing within the stripwise $\alpha$-coupling approach.

\section*{Acknowledgments}
The authors would like to thank Baptiste Arnould for his assistance with the CHAMPS Solver. This work was supported by the Natural Sciences and Engineering Research Council of Canada (NSERC), the Consortium de recherche et d’innovation en aérospatiale au Québec (CRIAQ), and CAE Inc. under the NSERC Alliance Grant (No. ALLRP 572207-22) and the CRIAQ project HAMAC. Calculations were performed on the Digital Research Alliance of Canada and Calcul Québec clusters. Lastly, we acknowledge the use of ChatGPT (GPT-5.2, 2026) for language polishing and editing.

\bibliography{sample}

\end{document}